\documentclass[draft,tightenlines,nofootinbib,preprint,aps,eqsecnum,amsmath,amssymb]{revtex4}

\begin{document}

\newcommand{\beq}{\begin{equation}}
\newcommand{\eeq}{\end{equation}}
\newcommand{\bea}{\begin{eqnarray}}
\newcommand{\eea}{\end{eqnarray}}
\newcommand{\cir}{{\buildrel \circ \over =}}

\newcommand{\sgn}{\mbox{\boldmath $\epsilon$}}

\newcommand{\on}{\stackrel{\circ}{=}}
\baselineskip 20pt

\title{Generalized Radar 4-Coordinates and Equal-Time Cauchy Surfaces
for Arbitrary Accelerated Observers.}

\author{David Alba}

\affiliation {Dipartimento di Fisica\\ Universita' di Firenze\\
Via G. Sansone 1\\ 50019 Sesto Fiorentino (FI), Italy\\ E-mail:
ALBA@FI.INFN.IT}

\author{Luca Lusanna}

\affiliation
 {Sezione INFN di Firenze\\ Via G. Sansone 1\\ 50019
Sesto Fiorentino (FI), Italy\\ E-mail: LUSANNA@FI.INFN.IT}

\begin{abstract}

All existing 4-coordinate systems centered on the world-line of an
accelerated observer are only locally defined like it happens for
Fermi coordinates both in special and general relativity. As a
consequence, it is not known how non-inertial observers can build
{\it equal-time surfaces} which a) correspond to a conventional
observer-dependent definition of synchronization of distant
clocks; b) are good Cauchy surfaces for Maxwell equations. Another
type of coordinate singularities generating the same problems are
those connected to the relativistic rotating coordinate systems
used in the treatment of the rotating disk and the Sagnac effect.
\medskip

We show that the use of Hamiltonian methods based on 3+1
splittings of space-time allows to define as many
observer-dependent globally defined radar 4-coordinate systems as
nice foliations of space-time with space-like hyper-surfaces
admissible according to M$\o$ller (for instance only
differentially  rotating relativistic coordinate system, but not
the rigidly rotating ones of non-relativistic physics, are
allowed). All these conventional notions of an {\it instantaneous
3-space} for an arbitrary observer can be empirically defined by
introducing generalizations of Einstein ${1\over 2}$ convention
for clock synchronization in inertial frames. Each admissible 3+1
splitting has two naturally associated congruences of time-like
observers: as a consequence every 3+1 splitting gives rise to
non-rigid non-inertial frames centered on anyone of these
observers. Only for the Eulerian observers the simultaneity leaves
are orthogonal to the observer world-line.

When there is a Lagrangian description of an isolated relativistic
system, its reformulation as a parametrized Minkowski theory
allows to show that all the admissible synchronization conventions
are {\it gauge equivalent}, as it also happens in canonical metric
and tetrad gravity, where, however, the chrono-geometrical
structure of space-time is dynamically determined.

The framework developed in this paper is not only useful for a
consistent description of the rotating disk, but is also needed
for the interpretation of the future ACES experiment on the
synchronization of laser cooled atomic clocks and for the
synchronization of the clocks on the three LISA spacecrafts.

\bigskip


\today

\end{abstract}

\maketitle

\newpage

\section{Introduction}

A physical observer is mathematically described by a
future-oriented time-like world-line $\gamma$, carrying an ideal
standard clock measuring proper time and  a tetrad field
\cite{1a}. Both in the Minkowski space-time of special relativity
and in the set of globally hyperbolic space-times, i.e. admitting
a global time function, compatible with Einstein general
relativity, the Lorentz signature of the 4-metric allows the
observer to identify the local light-cone in each point of
$\gamma$ and to speak of {\it events space-like with respect to a
point of $\gamma$}. In the case of special relativity an {\it
inertial} observer can use the clock moving along the
straight-line $\gamma$ to verify the validity of the two
independent postulates of the theory: in every inertial system the
{\it round-trip or two-ways} velocity of light A) is the same
($c$) and B) is isotropic.
\medskip

However, every physical observer has {\it no natural notion of
instantaneous 3-space} (of a {\it present} in colloquial terms) at
each point of $\gamma$ with all the clocks synchronized with the
one on $\gamma$ allowing the introduction of an associated notion
of {\it spatial distance} and an associated definition of {\it
one-way} velocity of light from $\gamma$ to every other time-like
world-line whose clock has been synchronized with the one on
$\gamma$.

\bigskip

In special relativity, usually, one considers only {\it global
rigid inertial reference frames} associated with an inertial
observer whose world-line $\gamma$ is a straight-line playing the
role of the time axis and with a point taken as origin of
Cartesian 4-coordinates $x^{\mu}$ for Minkowski space-time. By
means of an adiabatic slow transport of clocks, one identifies the
$x^o = const.$ space-like hyper-planes as the {\it instantaneous
3-space} with all the clocks synchronized \cite{2a}. The same
notion of {\it instantaneous 3-space} is arrived at if the
inertial observer $\gamma$ sends rays of light to another
time-like observer $\gamma_1$, who reflects them back towards
$\gamma$. Given the emission ($\tau_i$) and adsorption ($\tau_f$)
times on $\gamma$, the point $P$ of reflection on $\gamma_1$ is
assumed to be simultaneous with the point $Q$ on $\gamma$ where
$\tau_P\, {\buildrel {def}\over =}\, \tau_Q = \tau_i + {1\over
2}\, (\tau_f - \tau_i) = {1\over 2}\, (\tau_i + \tau_f)$. With
this so-called {\it Einstein's ${1\over 2}$ convention for the
synchronization of distant clocks} \cite{1} again the {\it
instantaneous 3-space} is the space-like hyper-plane $x^o =
const.$ {\it orthogonal to $\gamma$}, the point $Q$ is the
midpoint between the emission and adsorption points and, since
$\tau_P - \tau_i = \tau_f - \tau_P$, the {\it one-way} velocity of
light between $\gamma$ and every $\gamma_1$ is isotropic and equal
to the round-trip velocity of light $c$.

\medskip

The relativity principle then leads to the identification of the
kinematical Poincare' group as the set of transformations
connecting all the possible inertial observers with the appearance
of the standard effects of time dilation and length contraction.
As a consequence, the theoretical relevance of the inertial frames
seems to suggest that space-like hyper-planes orthogonal to
$\gamma$ are a natural notion of {\it instantaneous 3-space}.
However, as we shall see, there are many other geometrical
possibilities which simply do not correspond to rigid inertial
frames but to non-rigid non-inertial ones (the only ones existing
in general relativity due to a global interpretation of the
equivalence principle). Therefore, each observer has to {\it
stipulate some convention} defining a possible notion of
instantaneous 3-space and to study the transformation rules from a
convention to every other possible one. As a consequence, the
notion of instantaneous 3-space is both {\it observer-dependent
and conventional}.
\bigskip

Inertial frames are a limiting notion and, also disregarding
general relativity, all the observers on the Earth are non
inertial. According to the IAU 2000 Resolutions \cite{2}, for the
physics in the solar system one can consider the {\it Solar System
Barycentric Celestial Reference Frame} (with the axes identified
by fixed stars (quasars) of the Hypparcos catalog) as a
quasi-inertial frame. Instead the {\it Geocentric Celestial
Reference Frame}, with origin in the center of the geoid, is a
non-inertial frame whose axes are {\it non-rotating} with respect
to the Solar Frame. Every frame fixed on the surface of the Earth
is both non-inertial and rotating.

\medskip

Therefore, we need a definition of {\it an instantaneous 3-space}
for a non-inertial time-like observer, first in special relativity
and then in general relativity, and of what can be a non-rigid
non-inertial reference frame. In particular we have to define an
{\it accelerated coordinate system} and to face the problem of
rotating frames.

\medskip

Traditionally, given an arbitrary time-like observer with
world-line $\gamma$ and unit 4-velocity $u^{\mu}(\tau )$ ($\tau$
is the proper time measured by the clock of the observer), one
introduces the {\it Fermi coordinates} \cite{3,4,5} by considering
as {\it instantaneous 3-space} the space-like hyper-planes
orthogonal to the 4-velocity in each point of $\gamma$ \cite{3a}.
While the proper time $\tau$ labels the hyper-planes, the spatial
coordinates are defined by using three orthogonal 3-geodesics
emanating from $\gamma$ on the hyper-planes. However, this nice
local geometric construction defines 4-coordinates only locally in
a world-tube around $\gamma$, whose extension is determined by the
acceleration radii \cite{4a} where hyper-planes corresponding to
different values of $\tau$ intersect. Also Martzke-Wheeler
\cite{8} and Pauri-Vallisneri \cite{9} attempts, trying to
generalize Einstein ${1\over 2}$ convention to non-inertial
observers (and then to observers in general relativity), has the
same singularities of Fermi coordinates as shown in Ref.\cite{10}.
\medskip

To this type of coordinate-singularities we have to add the
singularities shown by all the rotating coordinate systems (the
problem of the {\it rotating disk}): the 4-metric expressed in
these coordinates has pathologies at the distance $R$ from the
rotation axis where $\omega\, R = c$ with $\omega$ being the
constant angular velocity of rotation \cite{5a}. Again, given the
unit 4-velocity field of the points of the rotating disk, there is
no notion of an instantaneous 3-space orthogonal to the associated
congruence of time-like observers, due to the non-zero vorticity
of the congruence \cite{6a}. Moreover,  an attempt to use Einstein
convention to synchronize the clocks on the rim of the disk fails
and one finds a synchronization gap (see Ref.\cite{14} and the
bibliography of Ref.\cite{15} for these problems and for the
Sagnac effect).

\bigskip

In conclusion this so-called {\it 1+3 point of view} (threading
splitting) of the accelerated observer is not able to build good
simultaneity hyper-surfaces, which, besides defining a
conventional instantaneous 3-space, are also good Cauchy surfaces
for Maxwell equations: only in this way we can find meaningful
solutions of these equations and to control the conservation laws.

\medskip

The problems quoted till now are not academic, but are becoming
relevant for measurements of one-way time transfer between Earth
and satellites with laser cooled high precision clocks \cite{16},
for future space navigation \cite{17}, for the Sagnac corrections
in the Global Positioning System \cite{18}, for the interpretation
of the future measurements of gravito-magnetism \cite{19,20} from
Gravity Probe B, for LISA \cite{21}.
\bigskip

In this paper we show that the {\it 3+1 point of view}  (slicing
splitting) used in the Hamiltonian treatment of dynamics, allows
to find a general solution of the previous problems.
\medskip

First of all, in Section II we shall remind which are the
admissible 4-coordinate transformations in Minkowski space-time
(frame-preserving diffeomorphisms) according to M$\o$ller
\cite{22}: with suitable restrictions at spatial infinity they
allow the definition of general notions of simultaneity as nice
foliations with space-like leaves, which replace the rigid
inertial reference frames with {\it non-rigid non-inertial
frames}. Note that for non-inertial frames there is no relativity
principle and no kinematical transformation group, like it happens
in general relativity, where the principle of general covariance
implies the replacement of the Poincare' group with the full
diffeomorphism group. In special relativity the frame-preserving
diffeomorphisms are playing the same role in replacing the
Poincare' group.

\medskip

Then we consider an arbitrary time-like observer with world-line
$\gamma$ ($x^{\mu}(\tau )$ are the coordinates in an inertial
frame and $\tau$ is the observer proper time), which intersects
the leaves of every admissible 3+1 splitting just in a point but
in general  is {\it not orthogonal} to any leaf. Given anyone of
the admissible 3+1 foliations, the world-line $\gamma$ is used to
build an {\it observer-dependent, globally defined, Lorentz-scalar
radar 4-coordinate system} by using the observer proper time
$\tau$ to label the leaves $\Sigma_{\tau}$ of the foliation and by
means of curvilinear 3-coordinates $\sigma^r$ \cite{7a}, with
origin on the world-line $\gamma$, on the leaves. The inverse of
the coordinate transformation $z^{\mu} \mapsto \sigma^A(z) =
(\tau, \vec \sigma )$, namely $\sigma^A \mapsto z^{\mu}(\tau ,\vec
\sigma )$, gives the embedding of the leaves $\Sigma_{\tau}$ in
Minkowski space-time as seen from the inertial frame. {\it The
hyper-surfaces $\Sigma_{\tau}$ are possible instantaneous
3-spaces, i.e. they are globally defined simultaneity
hyper-surfaces with all the clocks synchronized and also Cauchy
surfaces}.

\medskip

Let us remark that when we have a Lagrangian description of an
isolated relativistic system, we can arrive to a Lagrangian
depending also on the embedding $z^{\mu}(\tau ,\vec \sigma )$ and
being explicitly reparametrization invariant under
frame-preserving diffeomorphisms (the {\it parametrized Minkowski
theories} of Refs.\cite{23} and of the Appendix of Ref.\cite{24};
see Subsection C of Section II). Therefore, in this way we get a
{\it special relativistic notion of general covariance}, which is
a restriction of the general covariance under arbitrary
diffeomorphisms of general relativity to the frame-preserving
ones.

This implies that the transition from an admissible 3+1 splitting
to another one, i.e. the change of the convention of
synchronization of distant clocks and of instantaneous 3-space,
can be rephrased as a {\it gauge transformation}.

\medskip

Let us also note that many 3+1 splittings can be shown to agree
with the locality principle: {\it an accelerated observer at each
instant along its world-line is physically equivalent to an
otherwise identical momentarily comoving inertial observer},
namely a non-inertial observer passes through a continuous
infinity of hypothetical momentarily comoving inertial observers
\cite{8a}.

\medskip

As a byproduct of the study of M$\o$ller admissibility conditions
it will be shown that {\it global rigid rotations are not allowed}
in special (and also general) relativity: on  simultaneity and
Cauchy surfaces we must always have {\it differential rotations}.
Instead {\it global rigid translational accelerations are
allowed}. We also give the simplest set of rotating 4-coordinates
without pathologies, which can be used in the treatment of the
one-way time delay of signals between an Earth station and a
satellite (see Section VID of Ref.\cite{15}).

Moreover, in Appendix A, we solve the inverse problem of finding
admissible 3+1 splittings associated to a given unit 4-velocity
field with non-zero vorticity, like it happens with the rotating
disk. This allows to define genuine instantaneous 3-spaces with
synchronized clocks for a rotating disk, so that it is possible to
give a description of the Sagnac effect without synchronization
gap (see Sections VIB and C of Ref.\cite{15}).

\medskip

In conclusion the 3+1 point of view allows to find an infinite
number of admissible conventions for the definition of an
instantaneous 3-space, which can be used as a good Cauchy surface,
of an arbitrary accelerated observer. The pathologies of the Fermi
coordinates are avoided because the simultaneity surfaces, with
all the clocks synchronized, are {\it not hyper-planes orthogonal}
to the world-line of the observer.

\medskip

Moreover, in Section III we will show that the name radar
coordinates is justified, because they correspond to
generalizations of Einstein ${1\over 2}$ convention \cite{1,22}
\cite{9a}. In this Section we also outline the inverse problem of
how to build operationally this type of coordinates by using a
cluster of spacecrafts like the one used in the Global Positioning
System (GPS) \cite{18}.

\medskip

Then we make some concluding remarks regarding the extension of
these results to general relativity (see Ref.\cite{31} for the
status of the Hamiltonian formulation of metric and tetrad
gravity), where each solution of Einstein's equations dynamically
determines which 3+1 splittings of the globally hyperbolic
space-time can be associated to it, and some comments on which
problems have still to be clarified at the post-Newtonian level.

\bigskip

We refer to Ref.\cite{15} for an extended discussion and a rich
bibliography on all these problems, for a treatment of the
rotating disk and of the Sagnac effect  and for the determination
of the time-delay of signals from the Earth to a  satellite
(one-way velocity of light is required) within the 3+1 point of
view and, finally, for Maxwell equations in non-inertial frames.

\vfill\eject

\section{The 3+1 Point of View and Accelerated Observers. }

Let us consider the 3+1 splittings of Minkowski space-time
associated to its foliations with arbitrary space-like
hyper-surfaces and not only with space-like hyper-planes. Each of
these hyper-surfaces is both {\it a simultaneity surface and a
Cauchy surface} for the equations of motion of the relativistic
systems of interest. Having given a {\it notion of simultaneity},
there will be associated notions of {\it instantaneous 3-space},
{\it synchronization of distant clocks}, {\it spatial length} and
{\it one-way velocity of light}. After the choice of a foliation,
i.e. of a notion of simultaneity, we can determine, as we shall
see, which are the non-inertial observers compatible with that
notion of simultaneity.

\medskip

First of all we must find which 3+1 splittings of Minkowski
space-time are geometrically allowed as nice foliations whose
leaves are space-like hyper-surfaces. This leads to M$\o$ller
admissible coordinate transformations (Subsection A).

Then let us consider an arbitrary time-like observer whose
world-line $\gamma$ intersects each leaf of an admissible 3+1
splitting in a point. The world-line $\gamma$ can be used as a
centroid to define observer-adapted Lorentz-scalar radar
4-coordinates (Subsection B). As a consequence, each admissible
3+1 splitting {\it may} be chosen as a conventional notion of
instantaneous 3-space for an accelerated observer. In Subsection C
we show that every isolated relativistic system with a Lagrangian
description can be reformulated as a parametrized Minkowski theory
in which all the conventions are gauge equivalent.

In Subsection D we identify the 3+1 splittings which agree with
the locality hypothesis, namely in which it is evident that the
accelerated observer can be visualized as a sequence of comoving
inertial observers. Here we state which is the form assumed by
M$\o$ller admissibility conditions for such 3+1 splittings. As a
byproduct we show that, while we can have rigid non-inertial
frames with arbitrary translational acceleration, rigidly rotating
relativistic frames do not exist: globally defined rotating frames
must necessarily have differential rotations. In Subsection E we
identify the simplest global relativistic rotating frames as a
family of 3+1 splittings with space-like hyper-planes on which
there are differentially rotating coordinates.

\subsection{M$\o$ller Admissible Coordinates.}

Given an inertial system with Cartesian 4-coordinates $x^{\mu}$ in
Minkowski space-time and with the $x^o = const.$ simultaneity
hyper-planes, M$\o$ller, in Chapter VIII, Section 88 of
Ref.\cite{22} (see also Hilbert \cite{32} and Havas \cite{30}),
defines the {\it admissible coordinates transformations}
$x^{\mu}\, \mapsto\, y^{\mu} = f^{\mu}(x)$ [with inverse
transformation $y^{\mu}\, \mapsto\, x^{\mu} = h^{\mu}(y)$] as
those transformations whose associated metric tensor
$g_{\mu\nu}(y) = {{\partial h^{\alpha}(y)}\over {\partial
y^{\mu}}}\, {{\partial h^{\beta}(y)}\over {\partial y^{\nu}}}\,
\eta_{\alpha\beta}$ satisfies the following conditions (it can be
shown that the inverse metric $g^{\mu\nu}(y)$ satisfies the same
conditions)

\bea
 && \sgn\, g_{oo}(y) > 0,\nonumber \\
 &&{}\nonumber \\
 && \sgn\, g_{ii}(y) < 0,\qquad \begin{array}{|ll|} g_{ii}(y)
 & g_{ij}(y) \\ g_{ji}(y) & g_{jj}(y) \end{array}\, > 0, \qquad
 \sgn\, det\, [g_{ij}(y)]\, < 0,\nonumber \\
 &&{}\nonumber \\
 &&\Rightarrow det\, [g_{\mu\nu}(y)]\, < 0.
 \label{I1}
 \eea

These are the necessary and sufficient conditions for having
${{\partial h^{\mu}(y)}\over {\partial y^o}}$ behaving as the
velocity field of a relativistic fluid, whose integral curves, the
fluid flux lines, are the world-lines of time-like observers.
Eqs.(\ref{I1}) say:

i) the observers are time-like because $\sgn g_{oo} > 0$;

ii) that the hyper-surfaces $y^o = f^{o}(x) = const.$ are good
space-like simultaneity surfaces.

\medskip

Moreover we must ask that $g_{\mu\nu}(y)$ tends to a finite limit
at spatial infinity on each of the hyper-surfaces $y^o = f^{o}(x)
= const.$ If, like in the ADM canonical formulation of metric
gravity \cite{33,24}, we write $g_{oo} = \sgn\, (N^2 - g_{ij}\,
N^i\, N^j)$, $g_{oi} = g_{ij}\, N^j$ introducing the lapse ($N$)
and shift ($N^i$) functions, this requirement says that the lapse
function (i.e. the proper time interval between two nearby
simultaneity surfaces) and the shift functions (i.e. the
information about which points on two nearby simultaneity surfaces
are connected by the so-called {\it evolution vector field}
${{\partial h^{\mu}(y)}\over {\partial y^o}}$) must not diverge at
spatial infinity. This implies that at spatial infinity on each
simultaneity surface there is {\it no asymptotic either
translational or rotational acceleration} \cite{10a} and the
asymptotic line element is $ds^2 = g_{\mu\nu}(y)\, dy^{\mu}\,
dy^{\nu}\, \rightarrow_{spatial\, infinity}\, \sgn\, \Big(
F^2(y^o)\, (dy^o)^2 + 2\, G_i(y^o)\, dy^o\, dy^i - {d \vec
y}^2\Big)$. But this would break manifest covariance unless
$F(y^o) = 1$ and $G_i(y^o) = 0$. As a consequence, {\it the
simultaneity surfaces must tend to space-like hyper-planes at
spatial infinity}.

\bigskip

In this way all the admissible notions of simultaneity of special
relativity are formalized as 3+1 splittings of Minkowski
space-time by means of foliations whose leaves are space-like
hyper-surfaces tending to hyper-planes at spatial infinity. Let us
remark that admissible coordinate transformations $x^{\mu} \mapsto
y^{\mu} = f^{\mu}(x)$ constitute the most general extension of the
Poincare' transformations $x^{\mu} \mapsto y^{\mu} = a^{\mu} +
\Lambda^{\mu}{}_{\nu}\, x^{\nu}$ compatible with special
relativity. An important sub-group of M$\o$ller admissible
transformations  consists  the {\it frame-preserving}
diffeomorphisms: $x^o\, \mapsto\, y^o = f^o(x^o, \vec x)$, $\vec
x\, \mapsto\, \vec y = \vec f(\vec x)$, with inverse
transformations $x^o = h^o(y^o, \vec y)$, $\vec x = \vec h(\vec
y)$. Let us remark that the asymptotic conditions at spatial
infinity restrict M$\o$ller admissible transformations to those
which have the behavior ${{\partial h^o}\over {\partial y^o}}
\rightarrow 1$, ${{\partial h^o}\over {\partial y^i}} \rightarrow
0$, ${{\partial h^i}\over {\partial y^j}} \rightarrow \delta^i_j$
at spatial infinity.

\subsection{From M$\o$ller Admissible Coordinates to Radar
4-Coordinates adapted to an Arbitrary Accelerated Observer.}

It is then convenient to describe \cite{23,24} the simultaneity
surfaces of an admissible 3+1 splitting of Minkowski space-time
with {\it adapted Lorentz-scalar admissible coordinates}
$x^{\mu}\, \mapsto \sigma^A = (\tau ,\vec \sigma ) = f^A(x)$ [with
inverse $\sigma^A\, \mapsto\, x^{\mu} = z^{\mu}(\sigma ) =
z^{\mu}(\tau ,\vec \sigma )$] such that:

i) the scalar time coordinate $\tau$ labels the leaves
$\Sigma_{\tau}$ of the foliation ($\Sigma_{\tau} \approx R^3$);

ii) the scalar curvilinear 3-coordinates $\vec \sigma = \{
\sigma^r \}$ on each $\Sigma_{\tau}$ are defined with respect to
the world-line $\gamma$ of an arbitrary time-like centroid
$x^{\mu}(\tau )$ chosen as their origin;

iii) if $y^{\mu} = f^{\mu}(x)$ is any admissible coordinate
transformation describing the same foliation, i.e. if the leaves
$\Sigma_{\tau}$ are also described by $y^o = f^o(x) = const.$,
then, modulo reparametrizations, we must have $y^{\mu} =
f^{\mu}(z(\tau ,\vec \sigma )) = {\tilde f}^{\mu}(\tau ,\vec
\sigma ) = A^{\mu}{}_A\, \sigma^A$ with $A^o{}_{\tau} = const.$,
$A^o{}_r = 0$, so that we get $y^o = const.\, \tau$, $y^i =
A^i{}_A(\tau ,\vec \sigma )\, \sigma^A$. Therefore, modulo
reparametrizations, the $\tau$ and $\vec \sigma$ adapted
admissible coordinates are {\it intrinsic coordinates}, which are
mathematically allowed as charts in an enlarged atlas for
Minkowski space-time taking into account the extra structure of
the admissible 3+1 splittings. They are called {\it radar-like
4-coordinates} (see Section III for the justification of this
name) and, probably, they were introduced for the first time by
Bondi \cite{34}. The use of these Lorentz-scalar adapted
coordinates allows to make statements depending only on the
foliation but not on the 4-coordinates $y^{\mu}$ used for
Minkowski space-time.
\bigskip

{\it If we identify the centroid $x^{\mu}(\tau )$ with the
world-line $\gamma$ of an arbitrary time-like observer and $\tau$
with the observer proper time, we obtain as many globally defined
observer-dependent Lorentz-scalar radar 4-coordinates for an
accelerated observer as admissible 3+1 splittings of Minkowski
space-time and each 3+1 splitting can be viewed as a conventional
choice of an instantaneous 3-space and of a synchronization
prescription for distant clocks. The world-line $\gamma$ is not in
general orthogonal to the simultaneity leaves and Einstein
${1\over 2}$ convention is suitably generalized (see Section
III).}

\bigskip

The simultaneity hyper-surfaces $\Sigma_{\tau}$ are described by
their embedding $x^{\mu} = z^{\mu}(\tau ,\vec \sigma )$ in
Minkowski space-time [$(\tau ,\vec \sigma ) \mapsto z^{\mu}(\tau
,\vec \sigma )$, $R^3\, \mapsto \, \Sigma_{\tau} \subset M^4$] and
the induced metric is $g_{AB}(\tau ,\vec \sigma ) = z^{\mu}_A(\tau
,\vec \sigma )\, z^{\nu}_B(\tau ,\vec \sigma )\, \eta_{\mu\nu}$
with $z^{\mu}_A = \partial z^{\mu} / \partial \sigma^A$
\cite{11a}. Since the vector fields $z^{\mu}_r(\tau ,\vec \sigma
)$ are tangent to the surfaces $\Sigma_{\tau}$, the time-like
vector field of normals $l^{\mu}(\tau ,\vec \sigma )$ is
proportional to $\epsilon^{\mu}{}_{\alpha\beta\gamma}\,
z^{\alpha}_1(\tau ,\vec \sigma )\, z^{\beta}_2(\tau ,\vec \sigma
)\, z^{\gamma}_3(\tau ,\vec \sigma )$. Instead the time-like {\it
evolution vector field} is $z^{\mu}_{\tau}(\tau ,\vec \sigma ) =
N(\tau ,\vec \sigma )\, l^{\mu}(\tau ,\vec \sigma ) + N^r(\tau
,\vec \sigma )\, z^{\mu}_r(\tau ,\vec \sigma )$, so that we have
$dz^{\mu}(\tau ,\vec \sigma ) = z^{\mu}_{\tau}(\tau ,\vec \sigma
)\, d\tau + z^{\mu}_r(\tau ,\vec \sigma )\, d\sigma^r = N(\tau
,\vec \sigma )\, d\tau\, l^{\mu}(\tau ,\vec \sigma ) + [N^r(\tau
,\vec \sigma )\, d\tau + d\sigma^r]\, z^{\mu}_r(\tau ,\vec \sigma
)$.

Since the 3-surfaces $\Sigma_{\tau}$ are {\it equal time} 3-spaces
with all the clocks synchronized, the spatial distance between two
equal-time events will be $dl_{12} = \int^2_1 dl\,
\sqrt{{}^3g_{rs}(\tau ,\vec \sigma (l))\, {{d\sigma^r(l)}\over
{dl}}\, {{d\sigma^s(l)}\over {dl}}}\,\,$ ($\vec \sigma (l)$ is a
parametrization of the 3-geodesic $\gamma_{12}$ joining the two
events on $\Sigma_{\tau}$). Moreover, by using test rays of light
we can define the {\it one-way} velocity of light between events
on different $\Sigma_{\tau}$'s.

\medskip

Therefore the accelerated observer plus one admissible 3+1
splitting with the observer-dependent radar 4-coordinates define a
{\it non-rigid non-inertial reference frame} whose time axis is
the world-line $\gamma$ of the observer and whose instantaneous
3-spaces are the simultaneity hyper-surfaces $\Sigma_{\tau}$.

\bigskip

The main property of {\it each admissible foliation with
simultaneity surfaces} is that the embedding of the space-like
leaves of the foliation automatically determine two time-like
vector fields and therefore {\it two congruences of (in general)
non-inertial time-like observers} to be used to define
non-inertial frames with the given simultaneity surfaces:

i) The time-like vector field $l^{\mu}(\tau ,\vec \sigma )$ of the
normals to the simultaneity surfaces $\Sigma_{\tau}$ (by
construction surface-forming, i.e. irrotational), whose flux lines
are the world-lines of the so-called (in general non-inertial)
Eulerian observers. The simultaneity surfaces $\Sigma_{\tau}$ are
(in general non-flat) Riemannian 3-spaces in which the physical
system is visualized and in each point the tangent space to
$\Sigma_{\tau}$ is the local observer rest frame of the Eulerian
observer through that point. This 3+1 viewpoint is called {\it
hyper-surface 3+1 splitting}.

ii) The time-like evolution vector field $z^{\mu}_{\tau}(\tau
,\vec \sigma ) / \sqrt{\sgn\, g_{\tau\tau}(\tau ,\vec \sigma ) }$,
which in general is not surface-forming (i.e. it has non-zero
vorticity like in the case of the rotating disk). The observers
associated to its flux lines have the local observer rest frames,
the tangent 3-spaces orthogonal to the evolution vector field, not
tangent to $\Sigma_{\tau}$: there is no notion of 3-space for
these observers (1+3 point of view or {\it threading splitting})
and no visualization of the physical system in large. However
these observers can use the notion of simultaneity associated to
the embedding $z^{\mu}(\tau ,\vec \sigma )$, which determines
their 4-velocity. In this way we get non-inertial frames centered
on these observers, whose world-lines are {\it not orthogonal} to
the simultaneity surfaces. This 3+1 viewpoint is called {\it
slicing 3+1 splitting}.

\subsection{Parametrized Minkowski Theories.}

As said in the Introduction, when we have a Lagrangian description
of an isolated system, it can be extended to a  {\it parametrized
Minkowski theory} \cite{23,24}. In this approach, besides the
configuration variables of the isolated system, there are the
embeddings $z^{\mu}(\tau ,\vec \sigma )$ as extra {\it gauge
configuration} variables in a suitable Lagrangian determined in
the following way. Given the Lagrangian of the isolated system in
the Cartesian 4-coordinates of an inertial system, one makes the
coupling to an external gravitational field and then replaces the
external 4-metric with $g_{AB} = z^{\mu}_A\, \eta_{\mu\nu}\,
z^{\nu}_B$.

Therefore the resulting Lagrangian depends on the embedding
through the associated metric $g_{AB}$. It can be shown that, due
to the presence of {\it a special- relativistic type of general
covariance} (reparametrization invariance of the action under
frame-preserving diffeomorphisms), the transition from a foliation
to another one (i.e. a change of the notion of simultaneity) is a
{\it gauge transformation} of the theory. Therefore, in
parametrized Minkowski theories the {\it conventionalism of
simultaneity} is rephrased as a {\it gauge problem} (in a way
different from Refs.\cite{27,29}), i.e. as the {\it arbitrary
choice of a gauge fixing selecting a well defined notion of
simultaneity among those allowed by the gauge freedom}. Moreover,
for every isolated system there is a preferred notion of
simultaneity, namely the one associated with the 3+1 splitting
whose leaves are the {\it Wigner hyper-planes} orthogonal to the
conserved 4-momentum of the system: {\it this preferred
simultaneity, intrinsically selected by the isolated system,
identifies the intrinsic inertial rest frame centered on an
inertial observer having the system global 4-velocity and leads to
the Wigner-covariant rest-frame instant form of dynamics}
\cite{23}.

\bigskip

Besides scalar and spinning positive-energy particles,
Klein-Gordon, electro-magnetic, Yang-Mills and Dirac fields have
been reformulated as parametrized Minkowski theories \cite{23,24}.

\bigskip

This same state of affairs is also present in Hamiltonian metric
and tetrad gravity: the change of the notion of simultaneity is a
{\it gauge transformation} \cite{24,31,35,36}. The distinguishing
property of general relativity is that the solutions of Einstein's
equations determine {\it dynamically} which notions of
simultaneity are allowed \cite{36}.

\subsection{Admissible 4-Coordinates and the Locality Hypothesis:
Non-Existence of Rigid Rotating Reference Frames.}

Let us now consider a class of 4-coordinate transformations which
implements the idea of accelerated observers as sequences of
comoving observers  (the locality hypothesis) \cite{12a}  and let
us determine the conditions on the transformations to get a set of
admissible 4-coordinates. From now on we shall use Lorentz-scalar
radar-like 4-coordinates $\sigma^A = (\tau ; \vec \sigma )$
adapted to the foliation, whose simultaneity leaves are denoted
$\Sigma_{\tau}$.
\medskip

As we have said, the admissible embeddings $x^{\mu} = z^{\mu}(\tau
,\vec \sigma )$ [inverse transformations of $x^{\mu} \mapsto
\sigma^A(x)$], defined with respect to a given inertial system,
must tend to parallel space-like hyper-planes at spatial infinity.
If $l^{\mu} = l^{\mu}_{(\infty )} = \epsilon^{\mu}_{\tau}$,
$l^2_{(\infty )} = \sgn$, is the asymptotic normal, let us define
the asymptotic orthonormal tetrad $\epsilon^{\mu}_A$, $A=\tau
,1,2,3$, by using the standard Wigner boost for time-like
Poincare' orbits $L^{\mu}{}_{\nu}(l_{(\infty )}, {\buildrel \circ
\over l}_{(\infty )})$, ${\buildrel \circ \over l}_{(\infty )} =
(1; \vec 0)$: $\epsilon^{\mu}_A\, {\buildrel {def}\over =}\,
L^{\mu}{}_A(l_{(\infty )}, {\buildrel \circ \over l}_{(\infty
)})$, $\eta_{AB}=\epsilon^{\mu}_A\, \eta_{\mu\nu}\,
\epsilon^{\nu}_B$.
\medskip

Then a parametrization of the asymptotic hyper-planes is $z^{\mu}
= x^{\mu}_o + \epsilon^{\mu}_A\, \sigma^A = x^{\mu}(\tau ) +
\epsilon^{\mu}_r\, \sigma^r$ with $x^{\mu}(\tau ) = x^{\mu}_o +
\epsilon^{\mu}_{\tau}\, \tau$ a time-like straight-line (an
asymptotic inertial observer). Let us define a family of 3+1
splittings of Minkowski space-time by means of the following
embeddings \cite{13a}

\medskip

\bea
 z^{\mu}(\tau ,\vec \sigma ) &=& x^{\mu}_o +
\Lambda^{\mu}{}_{\nu}(\tau ,\vec \sigma )\, \epsilon^{\nu}_A\,
\sigma^A\, = {\tilde x}^{\mu}(\tau ) + F^{\mu}(\tau ,\vec \sigma
),\qquad F^{\mu}(\tau ,\vec 0) = 0,\nonumber \\
 &&{}\nonumber \\
 &&{\tilde x}^{\mu}(\tau ) = x^{\mu}_o + \Lambda^{\mu}{}_{\nu}(\tau ,\vec
 0)\, \epsilon^{\nu}_{\tau}\, \tau,\nonumber \\
 && F^{\mu}(\tau ,\vec
 \sigma ) = [\Lambda^{\mu}{}_{\nu}(\tau ,\vec \sigma ) -
 \Lambda^{\mu}{}_{\nu}(\tau ,\vec 0)]\, \epsilon^{\nu}_{\tau}\,
 \tau + \Lambda^{\mu}{}_{\nu}(\tau ,\vec \sigma )\,
 \epsilon^{\nu}_r\, \sigma^r,\nonumber \\
 &&{}\nonumber \\
 &&\Lambda^{\mu}{}_{\nu}(\tau ,\vec \sigma )\, {\rightarrow}_{|\vec
\sigma | \rightarrow \infty}\, \delta^{\mu}_{\nu},\quad
\Rightarrow\,\, z^{\mu}(\tau ,\vec \sigma )\, {\rightarrow}_{|\vec
\sigma | \rightarrow \infty}\,\, x^{\mu}_o + \epsilon^{\mu}_A\,
 \sigma^A = x^{\mu}(\tau ) + \epsilon^{\mu}_r\, \sigma^r,\nonumber \\
 &&{}
 \label{III3}
 \eea

\medskip

\noindent where $\Lambda^{\mu}{}_{\nu}(\tau ,\vec \sigma )$ are
Lorentz transformations ($\Lambda^{\mu}{}_{\alpha}\,
\eta_{\mu\nu}\, \Lambda^{\nu}{}_{\beta} = \eta_{\alpha\beta}$)
belonging to the component connected with the identity of
$SO(3,1)$. While the functions $F^{\mu}(\tau ,\vec \sigma )$
determine the form of the simultaneity surfaces $\Sigma_{\tau}$,
the centroid ${\tilde x}^{\mu}(\tau )$, corresponding to an {\it
arbitrary time-like observer} chosen as origin of the
3-coordinates on each $\Sigma_{\tau}$, determines how these
surfaces are packed in the foliation.

\bigskip

Since the asymptotic foliation with parallel hyper-planes, having
a constant vector field $l^{\mu} = \epsilon^{\mu}_{\tau}$ of
normals, defines an inertial reference frame, we see that the
foliation (\ref{III3}) with its associated non-inertial reference
frame is obtained from the asymptotic inertial frame by means of
{\it point-dependent Lorentz transformations}. As a consequence,
the integral lines, i.e. the non-inertial Eulerian observers  and
(non-rigid) non-inertial reference frames associated to this
special family of simultaneity notions, are parametrized as a
continuum of comoving inertial observers as required by the
locality hypothesis \cite{14a}.

\medskip

Let us remark that when an arbitrary isolated system is described
by a Minkowski parametrized theory, in which the embeddings
$z^{\mu}(\tau ,\vec \sigma )$ are {\it gauge} configuration
variables, the transition from the description of dynamics in one
of these non-inertial reference frames compatible with the
locality hypothesis to another arbitrary allowed reference frame,
like the one of footnote 10, is a {\it gauge transformation}: {\it
therefore in this case the locality hypothesis can always be
assumed valid modulo gauge transformations}.
\medskip

An equivalent parametrization of the embeddings of this family of
reference frames is
\medskip

\bea
 z^{\mu}(\tau ,\vec \sigma ) &=& x^{\mu}_o + \epsilon^{\mu}_B\,
\Lambda^B{}_A(\tau ,\vec \sigma )\, \sigma^A = x^{\mu}_o +
U^{\mu}_A(\tau ,\vec \sigma )\, \sigma^A
 = {\tilde x}^{\mu}(\tau ) + F^{\mu}(\tau ,\vec \sigma ),\nonumber \\
 &&{}\nonumber \\
 &&{\tilde x}^{\mu}(\tau ) = x^{\mu}_o +
U^{\mu}_{\tau}(\tau ,\vec 0)\, \tau ,\nonumber \\
 &&F^{\mu}(\tau ,\vec \sigma ) =
 [U^{\mu}_{\tau}(\tau ,\vec \sigma ) -
 U^{\mu}_{\tau}(\tau ,\vec 0)]\, \tau + U^{\mu}_r(\tau ,\vec
 \sigma )\, \sigma^r,
\label{III5} \eea

\noindent where we have defined:

 \bea
 \Lambda^B{}_A(\tau ,\vec \sigma ) &=& \epsilon^B_{\mu}\,
  \Lambda^{\mu}{}_{\nu}(\tau ,\vec \sigma )\,
\epsilon^{\nu}_A,\qquad  U^{\mu}_A(\tau ,\vec \sigma )\,
 \eta_{\mu\nu}\, U^{\nu}_B(\tau ,\vec \sigma ) =
 \epsilon^{\mu}_A\, \eta_{\mu\nu}\,
 \epsilon^{\nu}_B = \eta_{AB}, \nonumber \\
 &&{}\nonumber \\
 U^{\mu}_A(\tau ,\vec \sigma ) &=& \epsilon^{\mu}_B\,
 \Lambda^B{}_A(\tau ,\vec \sigma )\, {\rightarrow}_{|\vec \sigma |
  \rightarrow \infty}\, \epsilon^{\mu}_A,
 \label{III6}
 \eea

\noindent where $\epsilon^B_{\mu} = \eta_{\mu\nu}\, \eta^{BA}\,
\epsilon^{\nu}_A$ are the inverse tetrads.

\bigskip

A slight generalization of these embeddings allows to find
Nelson's \cite{37} 4-coordinate transformation (but extended from
$\vec \sigma$-independent Lorentz transformations
$\Lambda^{\mu}{}_{\nu} = \Lambda^{\mu}{}_{\nu}(\tau )$ to $\vec
\sigma$-dependent ones!) implying M$\o$ller rotating 4-metric
\cite{15a}

\bea
 z^{\mu}(\tau ,\vec \sigma ) &=& x^{\mu}_o + \epsilon^{\mu}_A\,
\Big[ \Lambda^A{}_B(\tau ,\vec \sigma )\, \sigma^B + V^A(\tau
,\vec \sigma )\Big],\nonumber \\
 &&{}\nonumber \\
 V^{\tau}(\tau ,\vec \sigma ) &=& \int^{\tau}_o d\tau_1\,
\Lambda^{\tau}{}_{\tau}(\tau_1 ,\vec \sigma ) -
\Lambda^{\tau}{}_{\tau}(\tau ,\vec \sigma )\, \tau,
 V^r(\tau ,\vec \sigma ) = \int^{\tau}_o d\tau_1\,
\Lambda^r{}_{\tau}(\tau_1, \vec \sigma ) - \Lambda^r{}_{\tau}(\tau
,\vec \sigma )\, \tau .\nonumber \\
 &&{}
 \label{III7}
 \eea

\noindent However, in general these 3+1 splittings do not produce
a metric satisfying Eqs.(\ref{I1}).

\bigskip

Let us study the conditions imposed by Eqs.(\ref{I1})  on the
foliations of the type (\ref{III5}) (for the others it is similar)
to find which ones correspond to admissible 4-coordinate
transformations. We shall represent each Lorentz matrix $\Lambda$
as the product of a Lorentz boost $B$ and a rotation matrix ${\cal
R}$ {\it to separate the translational from the rotational
effects} ($\vec \beta = \vec v/c$ are the boost parameters,
$\gamma (\vec \beta ) = 1/\sqrt{1- {\vec \beta}^2}$, ${\vec
\beta}^2 = (\gamma^2 - 1)/\gamma^2$, $B^{-1}(\vec \beta ) =
B(-\vec \beta )$; $\alpha$, $\beta$, $\gamma$ are three Euler
angles and $R^{-1} = R^T$)

\bea
 &&\Lambda (\tau ,\vec \sigma ) = B(\vec \beta (\tau ,\vec \sigma
 ))\, {\cal R}(\alpha (\tau ,\vec \sigma ), \beta (\tau ,\vec
 \sigma ), \gamma (\tau ,\vec \sigma )),\nonumber \\
 &&{}\nonumber \\
 &&B^A{}_B(\vec \beta ) = \left( \begin{array}{cc}
\gamma (\vec \beta )& \gamma (\vec \beta )\, \beta^s\\ \gamma
(\vec \beta )\, \beta^r& \delta^{rs} + {{\gamma^2(\vec \beta )\,
\beta^r\, \beta^s}\over {\gamma (\vec \beta ) + 1}}
\end{array} \right),\qquad
 {\cal R}^A{}_B(\alpha ,\beta ,\gamma ) = \left(
 \begin{array}{cc} 1 & 0\\ 0& R^r{}_s(\alpha ,\beta ,\gamma )
 \end{array} \right),\nonumber \\
 &&{}\nonumber \\
 &&R(\alpha ,\beta ,\gamma ) =\label{III8}\\
&&\nonumber\\
 &&= \left( \begin{array}{ccc}
\cos \alpha \cos \beta \cos \gamma -\sin \alpha \sin \gamma  &
\sin \alpha \cos \beta \cos \gamma +\cos \alpha \sin \gamma  &
-\sin \beta \cos \gamma
\\ -\cos \alpha \cos \beta \sin \gamma -\sin \alpha \cos \gamma  &
-\sin \alpha \cos \beta \sin \gamma +\cos \alpha \cos \gamma  &
\sin \beta \sin \gamma \\ \cos \alpha \sin \beta  & \sin \alpha
\sin \beta  & \cos \beta \end{array} \right).\nonumber
 \eea

Then we get

\begin{eqnarray*}
 z^{\mu}_{\tau}(\tau ,\vec \sigma ) &=& U^{\mu}_{\tau}(\tau ,\vec
 \sigma ) + \partial_{\tau}\, U^{\mu}_A(\tau ,\vec \sigma )\,
 \sigma^A =\nonumber \\
&&\nonumber\\
 &=& U^{\mu}_{\tau}(\tau ,\vec \sigma ) + U^{\mu}_B(\tau ,\vec \sigma
 )\, \Omega^B{}_A(\tau ,\vec \sigma )\, \sigma^A,
 \end{eqnarray*}

\begin{eqnarray*}
 z^{\mu}_r(\tau ,\vec \sigma ) &=& U^{\mu}_r(\tau ,\vec \sigma ) +
 \partial_r\, U^{\mu}_A(\tau ,\vec \sigma )\, \sigma^A =\nonumber \\
&&\nonumber\\
 &=& U^{\mu}_r(\tau ,\vec \sigma ) + U^{\mu}_B(\tau ,\vec \sigma
 )\, \Omega^B_{(r) A}(\tau ,\vec \sigma )\, \sigma^A,
 \end{eqnarray*}

\bea
 l^{\mu}(\tau ,\vec \sigma ) &=&
 {1\over {\sqrt{|\det\, g_{rs}(\tau ,\vec \sigma )|}}}\,
 \epsilon^{\mu}{}_{\alpha\beta\gamma}\, [ z^{\alpha}_1\,
 z^{\beta}_2\, z^{\gamma}_3](\tau ,\vec \sigma ),\nonumber \\
 &&{}\nonumber \\
  &&\mbox{ (normal to the simultaneity surfaces\, )},
 \label{III9}
 \eea

\noindent where we have introduced the following matrices

\begin{eqnarray*}
 &&\Omega^A{}_B = (\Lambda^{-1}\, \partial_{\tau}\, \Lambda
 )^A{}_B = ( {\cal R}^{-1}\, \partial_{\tau}\, {\cal R} + {\cal
 R}^{-1}\, B^{-1}\, \partial_{\tau}\, B\, {\cal R})^A{}_B
 =( \Omega_{\cal R} + {\cal R}^{-1}\, \Omega_B\, {\cal
 R})^A{}_B,\nonumber \\
 &&{}\nonumber \\
 &&\qquad \Omega_{\cal R} = {\cal R}^{-1}\, \partial_{\tau}\, {\cal R} =
 \left( \begin{array}{cc} 0& 0\\ 0& \Omega_R =
 R^{-1}\, \partial_{\tau}\, R \end{array} \right),
\end{eqnarray*}

\begin{eqnarray*}
 &&\qquad \Omega_B = B^{-1}(\vec \beta )\, \partial_{\tau}\, B(\vec \beta
 ) = - \partial_{\tau}\, B(-\vec \beta )\, B^{-1}(-\vec \beta ) =
 \nonumber \\
&&\nonumber\\
 &&\qquad = \left( \begin{array}{cc} 0& \gamma\, (\partial_{\tau}\beta^s +
 {{\gamma^2\, \vec \beta \cdot \partial_{\tau}\vec \beta\,
 \beta^s}\over {\gamma + 1}})\\ \gamma\, (\partial_{\tau}\beta^r +
 {{\gamma^2\, \vec \beta \cdot \partial_{\tau}\vec \beta\,
 \beta^r}\over {\gamma + 1}})& - {{\gamma^2}\over {\gamma + 1}}\,
 (\beta^r\, \partial_{\tau}\beta^s - \partial_{\tau}\beta^r\,
 \beta^s) \end{array} \right),\nonumber \\
 &&{}\nonumber \\
 &&{}\nonumber \\
 && \qquad \Omega = \left( \begin{array}{cc} 0&
\gamma\, (\partial_{\tau}\beta^u +
 {{\gamma^2\, \vec \beta \cdot \partial_{\tau}\vec \beta\,
 \beta^u}\over {\gamma + 1}})\, R^u{}_s\\
 R^{T r}{}_u\, \gamma\, (\partial_{\tau}\beta^u +
 {{\gamma^2\, \vec \beta \cdot \partial_{\tau}\vec \beta\,
 \beta^u}\over {\gamma + 1}})& \Omega^r_{R\, s} - {{\gamma^2}\over
 {\gamma + 1}}\, R^{T\, r}{}_u\, (\beta^u\, \partial_{\tau}\beta^v
 - \partial_{\tau}\beta^u\, \beta^v)\, R^v{}_s \end{array} \right),
 \end{eqnarray*}

\bea
 &&\Omega^A_{(r) B} = (\Lambda^{-1}\, \partial_r\, \Lambda )^A{}_B
  = ( {\cal R}^{-1}\, \partial_r\, {\cal R} + {\cal
 R}^{-1}\, B^{-1}\, \partial_r\, B\, {\cal R})^A{}_B
 =( \Omega_{{\cal R}\, (r)} + {\cal R}^{-1}\, \Omega_{B\, (r)}\, {\cal
 R})^A{}_B = \nonumber \\
&&\nonumber\\
 &&= \left( \begin{array}{cc} 0&
\gamma\, (\partial_r\beta^u +
 {{\gamma^2\, \vec \beta \cdot \partial_r\vec \beta\,
 \beta^u}\over {\gamma + 1}})\, R^u{}_s\\
 R^{T w}{}_u\, \gamma\, (\partial_r\beta^u +
 {{\gamma^2\, \vec \beta \cdot \partial_r\vec \beta\,
 \beta^u}\over {\gamma + 1}})& \Omega^w_{R\, s} - {{\gamma^2}\over
 {\gamma + 1}}\, R^{T\, w}{}_u\, (\beta^u\, \partial_r\beta^v
 - \partial_r\beta^u\, \beta^v)\, R^v{}_s \end{array}\right),
 \label{III10}
 \eea
\medskip

\noindent assumed to vanish at spatial infinity, $\Omega^A{}_B
(\tau ,\vec \sigma ), \Omega^A_{(r)\, B}(\tau ,\vec \sigma )\,
{\rightarrow}_{|\vec \sigma | \rightarrow \infty}\, 0$. {\it The
matrix $\Omega_B$  describes the translational velocity ($\vec
\beta$) and acceleration ($\partial_{\tau}\vec \beta$), while the
matrix $\Omega_R$ the rotational angular velocity}.

The $z^{\mu}_A$'s and the associated 4-metric are

\bea
 z^{\mu}_{\tau}(\tau ,\vec \sigma ) &=& \Big([1 + \Omega^{\tau}{}_r\,
 \sigma^r]\, U^{\mu}_{\tau} + \Omega^r{}_A\, \sigma^A\, U^{\mu}_r
 \Big)(\tau ,\vec \sigma ),\nonumber \\
&&\nonumber\\ z^{\mu}_r(\tau ,\vec \sigma ) &=&
\Big(\Omega^{\tau}_{(r)\, s}\, \sigma^s\,
 U^{\mu}_{\tau} + [\delta_r^s + \Omega^s_{(r)\, A}\, \sigma^A]\,
 U^{\mu}_s\Big)(\tau ,\vec \sigma ),
\label{III11} \eea

\noindent and

 \begin{eqnarray*}
 g_{\tau\tau}(\tau ,\vec \sigma ) &=& \Big(z^{\mu}_{\tau}\,
 \eta_{\mu\nu}\, z^{\nu}_{\tau}\Big)(\tau ,\vec \sigma
 )= \sgn\, \Big([ 1+ \Omega^{\tau}{}_r\, \sigma^r]^2 -
  \sum_r\, [\Omega^r{}_A\, \sigma^A]^2\Big)(\tau ,\vec \sigma ),\nonumber \\
&&\nonumber\\
 g_{r\tau}(\tau ,\vec \sigma ) &=& \Big( z^{\mu}_r\,
 \eta_{\mu\nu}\, z^{\nu}_{\tau}\Big)(\tau ,\vec \sigma )
 = \sgn\, \Big(\Omega^{\tau}_{(r)\, s}\, \sigma^s\, [1 + \Omega^{\tau}{}_u\, \sigma^u]
 -\nonumber \\
&&\nonumber\\
 &-& \sum_s\, \Omega^s{}_A\, \sigma^A\, [\delta^s_r + \Omega^s_{(r)\, A}\, \sigma^A]
 \Big)(\tau ,\vec \sigma ),
 \end{eqnarray*}

\bea
 g_{rs}(\tau ,\vec \sigma ) &=& \Big( z^{\mu}_r\, \eta_{\mu\nu}\,
 z^{\nu}_s\Big)(\tau ,\vec \sigma ) =
  \sgn\, \Big(- \delta_{rs} - [\Omega^r_{(s)\, A} + \Omega^s_{(r)\, A}]\,
 \sigma^A +\nonumber \\
&&\nonumber\\
 &+& \Omega^{\tau}_{(r)\, u}\, \Omega^{\tau}_{(s)\, v}\, \sigma^u\, \sigma^v -
 \sum_u\, \Omega^u_{(r)\, A}\, \Omega^u_{(s)\, A}\, \sigma^A\, \sigma^B
 \Big)(\tau ,\vec \sigma ).
 \label{III12}
 \eea

\medskip

Eqs.(\ref{I1}) are complicated restrictions on the parameters
$\vec \beta (\tau ,\vec \sigma )$, $\alpha (\tau ,\vec \sigma )$,
$\beta (\tau ,\vec \sigma )$, $\gamma (\tau ,\vec \sigma )$ of the
Lorentz transformations, which say that {\it translational
accelerations and rotational frequencies are not independent} but
must {\it balance each other} if Eqs.(\ref{III5}) describe the
inverse of an admissible 4-coordinate transformation.
\medskip

Let us consider two extreme cases.

\bigskip

A)  {\it Rigid non-inertial reference frames with translational
acceleration exist}. An example are the following embeddings,
which are compatible with the locality hypothesis only for $f(\tau
) = \tau$ (this corresponds to $\Lambda = B(\vec 0)\, {\cal
R}(0,0,0)$, i.e. to an inertial reference frame)

\medskip

\bea
 z^{\mu}(\tau ,\vec \sigma ) &=& x^{\mu}_o +
\epsilon^{\mu}_{\tau}\, f(\tau ) + \epsilon^{\mu}_r\,
\sigma^r,\nonumber \\
 &&{}\nonumber \\
 &&g_{\tau\tau}(\tau ,\vec \sigma ) = \sgn\,
 \Big({{d f(\tau )}\over {d\tau}}\Big)^2,\quad g_{\tau r}(\tau ,\vec \sigma )
 =0,\quad g_{rs}(\tau ,\vec \sigma ) = -\sgn\, \delta_{rs}.
 \label{III13}
 \eea

\medskip

This is a foliation with parallel hyper-planes with respect to a
centroid $x^{\mu}(\tau ) = x^{\mu}_o + \epsilon^{\mu}_{\tau}\,
f(\tau )$ (origin of 3-coordinates). The hyper-planes have
translational acceleration ${\ddot x}^{\mu}(\tau ) =
\epsilon^{\mu}_{\tau}\, \ddot f(\tau )$, so that they are not
uniformly distributed like in the inertial case $f(\tau ) = \tau$.

\bigskip

B) On the other hand  {\it rigid rotating reference frames do not
exist}. Let us consider the embedding (compatible with the
locality hypothesis) with $\Lambda = B(\vec 0)\, {\cal R}(\alpha
(\tau ),\beta (\tau ),\gamma (\tau ))$ and $x^{\mu}(\tau ) =
x^{\mu}_o + \epsilon^{\mu}_{\tau}\, \tau$
\medskip

\begin{eqnarray*}
 z^{\mu}(\tau ,\vec \sigma ) &=& x^{\mu}(\tau ) + \epsilon^{\mu}_r\,
R^r{}_s(\tau )\, \sigma^s,\nonumber \\
 &&{}\nonumber \\
 z^{\mu}_{\tau}(\tau ,\vec \sigma ) &=& {\dot x}^{\mu}(\tau ) +
 \epsilon^{\mu}_r\, {\dot R}^r{}_s(\tau )\, \sigma^s,\qquad
 z^{\mu}_r(\tau ) = \epsilon^{\mu}_s\, R^s{}_r(\tau ),\nonumber \\
 &&{}\nonumber \\
 g_{\tau\tau}(\tau ,\vec \sigma ) &=& \sgn\,
 \Big({\dot x}^2(\tau ) + 2\, {\dot x}_{\mu}(\tau )\,
 \epsilon^{\mu}_r\, {\dot R}^r{}_s(\tau )\, \sigma^s -
  \sgn \, {\dot R}^r{}_u(\tau )\, {\dot R}^r{}_v(\tau )\,
 \sigma^u\, \sigma^v\Big),
 \end{eqnarray*}

\bea
 g_{\tau r}(\tau ,\vec \sigma ) &=& \sgn\, \Big({\dot x}_{\mu}(\tau )\,
 \epsilon^{\mu}_s\, R^s{}_r(\tau ) - \epsilon\, {\dot R}^v{}_u(\tau )\,
 R^v{}_r(\tau )\, \sigma^u \Big),\nonumber\\
&&\nonumber\\
 g_{rs}(\tau ,\vec \sigma )
 &=& - \sgn\, R^u{}_r(\tau )\, R^u{}_s(\tau ),
 \label{III14}
 \eea
\medskip

\noindent which corresponds to a foliation with parallel
space-like hyper-planes with normal $l^{\mu} =
\epsilon^{\mu}_{\tau}$. It can be verified that it is not the
inverse of an admissible 4-coordinate transformation, because the
associated $g_{\tau\tau}(\tau ,\vec \sigma )$ has a zero at
\cite{16a}
\medskip

\beq
 \sigma = \sigma_R = {1\over {\Omega (\tau )}}\, \Big[- {\dot x}
_{\mu}(\tau )\, b^{\mu}_r(\tau )\, (\hat \sigma \times \hat
\Omega(\tau ))^r + \sqrt{{\dot x}^2(\tau ) + [{\dot x}_{\mu}(\tau
)\, b^{\mu}_r(\tau )\, (\hat \sigma \times \hat \Omega (\tau
))^2}]^2 \,\, \Big],
 \label{III15}
 \eeq

\noindent with $\quad \sigma_R \rightarrow \infty$ for $\Omega
\rightarrow 0$. At $\sigma = \sigma_R$ the time-like vector
$z^{\mu}_{\tau}(\tau ,\vec \sigma )$ becomes light-like (the {\it
horizon problem}), while for an admissible foliation with
space-like leaves it must always remain time-like.

\medskip

This pathology (the so-called horizon problem) is common to most
of the rotating coordinate systems (see Subsection D of the
Introduction of Ref.\cite{15} for a partial list of the existing
options). Let us remark that an analogous pathology happens on the
event horizon of the Schwarzschild black hole. Also in this case
we have a coordinate singularity where the time-like Killing
vector of the static space-time becomes light-like. For the
rotating Kerr black hole the same coordinate singularity happens
already at the boundary of the ergosphere \cite{38}. Also in the
existing theory of rotating relativistic stars \cite{39}, where
differential rotations are replacing the rigid ones in model
building, it is assumed that in certain rotation regimes an
ergosphere may form \cite{40}: again, if one uses 4-coordinates
adapted to the Killing vectors, one gets a similar coordinate
singularity.

\bigskip

In the next Subsection we shall consider the minimal modification
of Eq.(\ref{III14}) so to obtain the inverse of an allowed
4-coordinate transformation to a rotating coordinate system.

\subsection{The Simplest Notion of Simultaneity when Rotations are
Present.}

Let us look for the simplest embedding $x^{\mu} = z^{\mu}(\tau
,\vec \sigma )$, inverse of an admissible 4-coordinate
transformation $x^{\mu} \mapsto \sigma^A$ compatible with the
locality hypothesis, which contains a rotating reference frame,
with also translational acceleration, of the type of
Eq.(\ref{III14}). The minimal modification of Eq.(\ref{III14}) is
to replace the rotation matrix $R(\tau )$ with $R(\tau ,|\vec
\sigma|)$, namely the rotation varies as a function of some radial
distance $|\vec \sigma |$ ({\it differential rotation}) from the
arbitrary time-like world-line $x^{\mu}(\tau )$, origin of the
3-coordinates on the simultaneity surfaces. Since the
3-coordinates $\sigma^r$ are Lorentz scalar we shall use the
radial distance $\sigma = |\vec \sigma| = \sqrt{\delta_{rs}\,
\sigma^r\, \sigma^s }$, so that $\sigma^r = \sigma \, {\hat
\sigma}^r$ with $\delta_{rs}\, {\hat \sigma}^r\, {\hat \sigma}^s =
1$. Therefore let us replace Eq.(\ref{III14}) with the following
embedding

\medskip

\bea
 &&z^{\mu}(\tau ,\vec \sigma ) = x^{\mu}(\tau ) + \epsilon^{\mu}_r\,
R^r{}_s(\tau , \sigma )\, \sigma^s\,
 {\buildrel {def}\over =}\, x^{\mu}(\tau ) + b^{\mu}_r(\tau
 ,\sigma )\, \sigma^r,\nonumber \\
 &&{}\nonumber \\
 &&R^r{}_s(\tau ,\sigma ) {\rightarrow}_{\sigma \rightarrow
 \infty} \delta^r_s,\qquad \partial_A\, R^r{}_s(\tau
 ,\sigma )\, {\rightarrow}_{\sigma \rightarrow
 \infty}\, 0,\nonumber \\
 &&{}\nonumber \\
 &&b^{\mu}_s(\tau ,\sigma ) = \epsilon^{\mu}_r\, R^r{}_s(\tau
 ,\sigma )\, {\rightarrow}_{\sigma \rightarrow
 \infty}\, \epsilon^{\mu}_s,\quad [b^{\mu}_r\, \eta_{\mu\nu}\, b^{\nu}_s](\tau ,\sigma )
 = - \sgn\, \delta_{rs}.
 \label{IV1}
 \eea
\medskip

Since $z^{\mu}_r(\tau ,\vec \sigma ) = \epsilon^{\mu}_s\,
\partial_r\, [R^s{}_u(\tau ,\sigma )\, \sigma^u]$, it follows that
the normal to the simultaneity surfaces is $l^{\mu} =
\epsilon^{\mu}_{\tau}$, namely the hyper-surfaces are {\it
parallel space-like hyper-planes}. These hyper-planes have
translational acceleration ${\ddot x}^{\mu}(\tau )$ and a rotating
3-coordinate system with rotational frequency

\begin{eqnarray*}
 \Omega^r(\tau ,\sigma ) &=& - {1\over 2}\, \epsilon^{ruv}\, \Big[
 R^{-1}(\tau ,\sigma )\, {{\partial R(\tau ,\sigma )}\over
 {\partial \tau}}\Big]^{uv}\, {\rightarrow}_{\sigma \rightarrow
 \infty}\, 0,\nonumber \\
 &&{}\nonumber \\
 &&\Downarrow \nonumber \\
 &&{}\nonumber \\
 {{\partial b^{\mu}_s(\tau ,\sigma )}\over {\partial \tau}} &=&
 \epsilon^{\mu}_r\, {{\partial R^r{}_s(\tau ,\sigma )}\over
 {\partial \tau}} = - \epsilon^{suv}\, \Omega^u(\tau ,\sigma )\,
 b^{\mu}_v(\tau ,\sigma ),
 \end{eqnarray*}

\bea
 \Omega^1(\tau ,\sigma ) &=&\Big[ \partial_{\tau} \beta\, \sin\,
 \gamma - \partial_{\tau} \alpha\, \sin\, \beta\, \cos\,
 \gamma\Big](\tau ,\sigma ),\nonumber \\
&&\nonumber\\
 \Omega^2(\tau ,\sigma ) &=& \Big[\partial_{\tau} \beta\, \cos\,
 \gamma + \partial_{\tau} \alpha\, \sin\, \beta\, \sin\,
 \gamma\Big](\tau ,\sigma ),\nonumber \\
&&\nonumber\\
 \Omega^3(\tau ,\sigma ) &=& \Big[ \partial_{\tau} \gamma +
 \partial_{\tau} \alpha\, \cos\, \beta \Big](\tau ,\sigma ).
  \label{IV2}
  \eea

\noindent In the last three lines we used Eqs.(\ref{III8}) to find
the angular velocities. Moreover we can define

 \bea
&&\Omega_{(r)}(\tau ,\sigma ) = \Big[ R^{-1}\, \partial_r\,
 R\Big](\tau ,\sigma ) = 2{\hat \sigma}^r\, \Big[R^{-1}\,
 {{\partial R}\over {\partial \sigma }}\Big](\tau ,\sigma )\,
 {\rightarrow}_{\sigma \rightarrow \infty}\, 0,\nonumber \\
&&\nonumber\\
 &&\Omega^u_{(r) v}(\tau ,\sigma )\, \sigma^v = \Phi_{uv}(\tau
 ,\sigma )\, {{\sigma^r\, \sigma^v}\over {\sigma}},\qquad
 \Phi_{uv} = - \Phi_{vu},
\label{IV3} \eea

\medskip

As a consequence we have

\bea
 {\dot x}^{\mu}(\tau ) &=& \sgn\, \Big([{\dot x}_{\nu}(\tau
 )\, l^{\nu}]\, l^{\mu} - \sum_r\, [{\dot x}_{\nu}(\tau )\,
 \epsilon^{\nu}_r]\, \epsilon^{\mu}_r \Big),\nonumber \\
 &&{}\nonumber \\
 z^{\mu}_{\tau}(\tau ,\vec \sigma ) &=& N(\tau ,\vec \sigma )\,
 l^{\mu} + N^r(\tau ,\vec \sigma )\, z^{\mu}_r(\tau ,\vec \sigma )
 =\nonumber \\
&&\nonumber\\
 &=& {\dot x}^{\mu}(\tau ) - \epsilon^{suv}\, \Omega^u(\tau
 ,\sigma )\, b^{\mu}_v(\tau ,\sigma )\, \sigma^s =\nonumber \\
&&\nonumber\\
 &=& {\dot x}^{\mu}(\tau ) - (\vec \sigma \times \vec \Omega (\tau ,\sigma ))^r\,
 b^{\mu}_r(\tau ,\sigma )\, {\rightarrow}_{\sigma \rightarrow
 \infty}\, {\dot x}^{\mu}(\tau ),\nonumber \\
&&{}\nonumber \\
 z^{\mu}_r(\tau ,\vec \sigma ) &=& \epsilon^{\mu}_s\, \Big[
 R^s{}_r(\tau ,\sigma ) + \partial_r\, R^s{}_u(\tau ,\sigma )\,
 \sigma^u\Big] =\nonumber \\
&&\nonumber\\
 &=& b^{\mu}_s(\tau ,\sigma )\, \Big[ \delta^s_r + \Omega^s_{(r)
 u}(\tau ,\sigma )\, \sigma^u\Big]\, {\rightarrow}_{\sigma \rightarrow
 \infty}\, \epsilon^{\mu}_r,
\label{IV4} \eea

\noindent and then we obtain

 \bea
 g_{\tau\tau}(\tau ,\vec \sigma ) &=& {\dot x}^2(\tau ) - 2\,
 {\dot x}_{\mu}(\tau )\, b^{\mu}_r(\tau ,\sigma )\, (\vec \sigma
 \times \vec \Omega (\tau ,\sigma ))^r - \sgn\,
(\vec{\sigma}\times\vec{\Omega})^2
=\nonumber \\
&&\nonumber\\
 &=& \Big[ N^2 - g_{rs}\, N^r\, N^s\Big](\tau ,\vec
 \sigma ),\nonumber \\
&&\nonumber\\
 g_{\tau r}(\tau ,\vec \sigma ) &=& \Big[ g_{rs}\, N^s\Big](\tau ,\vec \sigma )
 = \nonumber\\
&&\nonumber\\
&=&{\dot x}_{\mu}(\tau )\,
 b^{\mu}_r(\tau ,\sigma )\, \Big[ \delta^v_r + \Omega^v_{(r)
 u}(\tau ,\sigma )\, \sigma ^u\Big] + \sgn\,
 [\vec \sigma \times \vec \Omega (\tau ,\sigma
 )]^s\, \Big[ \delta^s_r + \Omega^s_{(r) u}(\tau ,\sigma )\,
 \sigma^u\Big],\nonumber \\
&&\nonumber\\
 - \sgn\,g_{rs}(\tau ,\vec \sigma ) &=&  \delta_{rs} +
 \Big(\Omega^r_{(s) u}(\tau ,\sigma ) + \Omega^s_{(r) u}(\tau
 ,\sigma )\Big)\, \sigma^u
 + \sum_w\, \Omega^w_{(r) u}(\tau ,\sigma )\, \Omega^w_{(s)
 v}(\tau ,\sigma )\, \sigma^u\, \sigma^v.\nonumber \\
 &&{}
 \label{IV5}
 \eea

 The requirement that $g_{\tau\tau}(\tau ,\vec \sigma )$ and
 $g_{\tau r}(\tau ,\vec \sigma )$ tend to finite limits at spatial
 infinity puts the restrictions

 \begin{eqnarray*}
 | \vec \Omega (\tau ,\sigma )|,\,&& |\Omega^u_{(r) v}(\tau
 ,\sigma )|\,\, {\rightarrow}_{\sigma \rightarrow \infty}\,
 O(\sigma^{-(1+\eta )}),\quad \eta > 0,\nonumber \\
 &&{}\nonumber \\
 &&\Downarrow \nonumber \\
 &&{}\nonumber \\
 \partial_A\, R^r{}_s(\tau ,\sigma ) &{\rightarrow}_{\sigma \rightarrow
 \infty}& O(\sigma^{-(1+\eta )}),\, \Rightarrow\, R^r{}_s(\tau
 ,\sigma )\, {\rightarrow}_{\sigma \rightarrow \infty}\,
 O(\sigma^{-(1+\eta )}),\nonumber \\
 &&{}\nonumber \\
 z^{\mu}_{\tau}(\tau ,\vec \sigma ) &{\rightarrow}_{\sigma \rightarrow
 \infty}& {\dot x}^{\mu}(\tau ) + O(\sigma^{-\eta}),\nonumber \\
&&\nonumber\\
 b^{\mu}_r(\tau ,\sigma ) &{\rightarrow}_{\sigma \rightarrow
 \infty}& \epsilon^{\mu}_r + O(\sigma^{-(1+\eta )}),\qquad
 z^{\mu}_r(\tau ,\vec \sigma )\, {\rightarrow}_{\sigma \rightarrow \infty}\,
 \epsilon^{\mu}_r + O(\sigma^{-(1+\eta )}),\nonumber \\
&&\nonumber\\
 N^r(\tau ,\vec \sigma )\, z^{\mu}_r(\tau ,\vec \sigma )\,
 &{\rightarrow}_{\sigma \rightarrow \infty}&\, - \sgn\, {\dot
 x}_{\nu}(\tau )\, \epsilon^{\nu}_r\, \epsilon^{\mu}_r +
 O(\sigma^{-(1+2\eta )}),\nonumber \\
&&\nonumber\\
 N^r(\tau ,\vec \sigma ) &{\rightarrow}_{\sigma \rightarrow
 \infty}& -\sgn\, \delta^{rs}\, {\dot x}_{\nu}(\tau )\,
 \epsilon^{\nu}_s + O(\sigma^{-\eta}),
 \end{eqnarray*}

\bea
 N(\tau ,\vec \sigma )&& l^{\mu} = [z^{\mu}_{\tau} - N^r\,
 z^{\mu}_r](\tau ,\vec \sigma )\, {\rightarrow}_{\sigma
 \rightarrow \infty}\, \sgn\, [{\dot x}_{\nu}(\tau )\,
 l^{\nu}]\, l^{\mu} + O(\sigma^{-\eta}),\nonumber \\
 &&{}\nonumber \\
 g_{\tau\tau}(\tau ,\vec \sigma ) &{\rightarrow}_{\sigma \rightarrow
 \infty}& {\dot x}^2(\tau ) + O(\sigma^{-2\eta}),\nonumber \\
&&\nonumber\\
 g_{\tau r}(\tau ,\vec \sigma ) &{\rightarrow}_{\sigma \rightarrow
 \infty}& {\dot x}_{\mu}(\tau )\, \epsilon^{\mu}_r +
 O(\sigma^{-\eta}),\nonumber \\
&&\nonumber\\
 g_{rs}(\tau ,\vec \sigma ) &{\rightarrow}_{\sigma \rightarrow
 \infty}& - \sgn\, \delta_{rs} + O(\sigma^{-\eta}).
  \label{IV6}
  \eea

\bigskip

Let us look for a family of rotation matrices $R^r{}_s(\tau
,\sigma )$ satisfying the condition $\sgn\, g_{\tau\tau}(\tau
,\vec \sigma ) > 0$ of Eqs.(\ref{I1}).
\medskip

Let us make the ansatz that the Euler angles of $R(\alpha ,\beta
,\gamma )$ have the following factorized dependence on $\tau$ and
$\sigma$

\beq
 \alpha (\tau ,\sigma) =F(\sigma )\, \tilde \alpha (\tau
 ),\qquad
 \beta (\tau ,\sigma ) = F(\sigma )\, \tilde \beta (\tau
 ),\qquad
 \gamma (\tau ,\sigma )=F(\sigma )\, \tilde \gamma (\tau
 ),
 \label{IV7}
 \eeq

\noindent with

 \beq
  F(\sigma ) > 0,\quad {{d F(\sigma )}\over {d
\sigma }}\not= 0,\qquad F(\sigma )\, {\rightarrow}_{\sigma
\rightarrow \infty}\,
 O(\sigma^{-(1+\eta )}).
 \label{IV8}
 \eeq

We get

 \bea
  \Omega^1(\tau ,\sigma ) &=& F(\sigma )\, \Big( {\dot {\tilde
 \beta}}(\tau )\, \sin\, [F(\sigma )\, \tilde \gamma (\tau )] -
 {\dot {\tilde \alpha}}(\tau )\, \sin\, [F(\sigma )\, \tilde \beta
 (\tau )]\, \cos\, [F(\sigma )\, \tilde \gamma (\tau
 )]\Big),\nonumber \\
&&\nonumber\\
 \Omega^2(\tau ,\sigma ) &=& F(\sigma )\, \Big( {\dot {\tilde
 \beta}}(\tau )\, \cos\, [F(\sigma )\, \tilde \gamma (\tau )] +
 {\dot {\tilde \alpha}}(\tau )\, \sin\, [F(\sigma )\, \tilde \beta
 (\tau )]\, \sin\, [F(\sigma )\, \tilde \gamma (\tau
 )]\Big),\nonumber \\
&&\nonumber\\
 \Omega^3(\tau ,\sigma ) &=& F(\sigma )\, \Big( {\dot {\tilde
 \gamma}}(\tau ) + {\dot {\tilde \alpha}}(\tau )\, \cos\, [F(\sigma
 )\, \tilde \beta (\tau )]\Big),\nonumber \\
 &&{}\nonumber \\
 &&\Downarrow \nonumber \\
 &&{}\nonumber \\
 \Omega^r(\tau ,\sigma ) &=& F(\sigma )\, \tilde \Omega (\tau
 ,\sigma )\, {\hat n}^r(\tau ,\sigma ),\qquad {\hat n}^2(\tau
 ,\sigma ) = 1,\nonumber \\
 &&{}\nonumber \\
 0 &<& \tilde \Omega (\tau ,\sigma )\, \leq\, 2\, \max\, \Big( {\dot
 {\tilde \alpha}}(\tau ), {\dot {\tilde \beta}}(\tau ), {\dot
 {\tilde \gamma}}(\tau ) \Big) = 2\, M_1.
 \label{IV9}
 \eea
\medskip

Since $l^{\mu} = \epsilon^{\mu}_{\tau}\, {\buildrel {def}\over
=}\, b^{\mu}_{\tau}$ and $b^{\mu}_r(\tau ,\sigma )$ form an
orthonormal tetrad [$b^{\mu}_A(\tau ,\sigma )\, \eta_{\mu\nu}\,
b^{\nu}_B(\tau ,\sigma ) = \eta_{AB}$], let us decompose the
future time-like 4-velocity ${\dot x}^{\mu}(\tau )$ on it
($v_l(\tau )$ is the asymptotic lapse function)

\bea
 {\dot x}^{\mu}(\tau ) &=& v_l(\tau )\, l^{\mu} -
 \sum_r\, v_r(\tau ,\sigma )\, b^{\mu}_r(\tau ,\sigma
 )\nonumber \\
 &&{}\nonumber \\
 v_l(\tau ) &=& \sgn\, {\dot x}_{\mu}(\tau )\, l^{\mu} > 0,\qquad
 v_r(\tau ,\sigma ) = \sgn\, {\dot x}_{\mu}(\tau )\, b^{\mu}_r(\tau
 ,\sigma ),\nonumber \\
 &&{}\nonumber \\
 \sgn\, {\dot x}^2(\tau ) &=& v^2_l(\tau ) - \sum_r\,
 v^2_r(\tau ,\sigma ) > 0,\, \Rightarrow \sum_r\, v^2_r(\tau
 ,\sigma ) = {\vec v}^2(\tau ,\sigma ) \equiv {\vec v}^2(\tau ) <
 v^2_l(\tau ),
\label{IV10} \eea

We add the condition

 \beq
  |\vec v(\tau )| \leq {{v_l(\tau )}\over
K},\qquad K > 1.
 \label{IV11}
 \eeq

This condition is slightly stronger than the last of
Eqs.(\ref{IV10}), which  does not exclude the possibility that the
observer in $\vec{\sigma}=0$ has a time-like 4-velocity ${\dot
x}^{\mu}(\tau )$ which, however, becomes light-like at $\tau = \pm
\infty$ \cite{17a}. The condition (\ref{IV11}) excludes this
possibility. In other words the condition (\ref{IV11}) tell us
that the observer is without event-horizon, namely he can explore
all the Minkowski space-time by light-signal.

\medskip

Then the condition $\sgn\, g_{\tau\tau}(\tau ,\vec \sigma ) > 0$
becomes

\bea
 && \sgn\, g_{\tau\tau}(\tau ,\vec \sigma ) =\nonumber\\
&&\nonumber\\ &=& \sgn\, {\dot
 x}^2(\tau ) - 2\,  \sigma\, F(\sigma )\, \tilde \Omega
 (\tau ,\sigma )\, \sum_r\, v_r(\tau ,\sigma )\, \Big[\hat \sigma
 \times \hat n(\tau ,\sigma )\Big]^r
-\sigma^2\, {\tilde \Omega}^2(\tau ,\sigma )\, F^2(\sigma )\,
 \Big[\hat \sigma \times \hat n(\tau ,\sigma )\Big]^2 =\nonumber\\
&&\nonumber\\
 &=& c^2(\tau ) - 2\, b(\tau ,\vec \sigma )\, X(\tau ,\sigma ) -
 a^2(\tau ,\vec \sigma )\, X^2(\tau ,\sigma ) > 0
\label{IV12}
 \eea

\noindent where we have defined

 \bea
 &&c^2(\tau ) =  \sgn\, {\dot x}^2(\tau ) = v^2_l(\tau ) -
 {\vec v}^2(\tau ) > 0,\qquad c^2(\tau ) \geq {{K^2-1}\over
 {K^2}}\, v^2_l(\tau ),\nonumber \\
 &&{}\nonumber \\
 &&b(\tau ,\vec \sigma ) = \sum_r\, v_r(\tau ,\sigma
 )\, \Big[ \hat \sigma \times \hat n(\tau ,\sigma
 )\Big]^r,\nonumber \\
&&\nonumber\\
  && |b(\tau ,\vec \sigma )| \leq |\vec v(\tau )| < v_l(\tau
  ),\,\mbox{ or }\, |b(\tau ,\vec \sigma )| \leq {{v_l(\tau )}\over
  K},\,\,  K > 1,\nonumber \\
  &&{}\nonumber \\
  &&a^2(\tau ,\vec \sigma ) =\Big[\hat \sigma \times \hat n(\tau
  ,\sigma )\Big]^2 > 0,\quad a^2(\tau ,\vec \sigma ) \leq 1,\quad
  b^2(\tau ,\vec \sigma ) + a^2(\tau ,\sigma )\, c^2(\tau ) >
  0,\nonumber \\
  &&{}\nonumber \\
  &&X(\tau ,\sigma ) = \sigma\, F(\sigma )\, \tilde \Omega (\tau
  ,\sigma ).
 \label{IV13}
 \eea

\medskip

The study of the equation $a^2\, X^2 + 2\, b\, X - c^2 = A^2\, (X
- X_{+})\, (X - X_{-}) = 0$, with solutions $X_{\pm} = {1\over
{a^2}}\, (- b \pm \sqrt{b^2 + a^2\, c^2})$, shows that $\sgn\,
g_{\tau\tau} > 0$ implies $X_{-} < X < X_{+}$; being $X_-<0$ and
$X>0$ [see Eq.(\ref{IV9})], we have that a half of the conditions
($X_-<X$) is always satisfied. We have only to discuss the
condition $X<X_+$.
\medskip

Since $- v_{l}/K \leq b \leq v_l/K$, when $b$ increases in this
interval $X_{+}$ decrease with $b$. This implies
\[
X_{+} > {1\over {a^2}}\, \Big(- {{v_l}\over K} +
\sqrt{{{v^2_l}\over {K^2}} + a^2\, c^2 }\Big),
\]
so that $c^2 \geq {{K^2 - 1}\over {K^2}}\, v^2_l$ implies that we
will have $g_{\tau\tau} > 0$ if $0< X < {{v_l}\over {K\, a^2}}\,
(\sqrt{1 + (K^2 - 1)\, a^2} - 1)$, namely if the function
$F(\sigma )$ satisfies the condition

\[
 | F(\sigma )| <
{{v_l(\tau )}\over {K\, \sigma\, a^2(\tau ,\vec \sigma )\,
 \tilde \Omega (\tau ,\sigma )}}\,
\Big(\sqrt{1 + (K^2 - 1)\, a^2(\tau ,\vec \sigma )} - 1\Big)=
{{v_l(\tau )}\over {K}\,\tilde \Omega (\tau ,\sigma )}\,g(a^2).
\]

\medskip
Since $a^2\leq 1$ and $g(x)=(1/x)(\sqrt{1+(K^2-1)x}-1)$ is
decreasing for $x$ increasing in the interval $0<x<1$ ($K>1$), we
get $g(a^2)>g(1)=K-1$ and the stronger condition
\[
| F(\sigma )| < {{v_l(\tau )}\over {K\,\tilde \Omega (\tau ,\sigma
)}}\,(K-1).
\]

\medskip

The condition (\ref{IV9}) on the Euler angles and the fact that
 Eq.(\ref{IV11}) implies $\min v_l(\tau )=m>0$  lead to the
 following final condition on $F(\sigma)$ \cite{18a}

\bea
 &&0< F(\sigma ) <
 {m\over {2\, K\, M_1\, \sigma}}\,(K-1)=\frac{1}{M\,\sigma},\qquad
 {{d F(\sigma )}\over {d \sigma }} \not= 0,\nonumber \\
 &&{}\nonumber \\
 \mbox{ or }\qquad&&| \partial_{\tau} \alpha (\tau ,\sigma )|,
  | \partial_{\tau} \beta (\tau ,\sigma )|,
   | \partial_{\tau} \gamma (\tau ,\sigma )| <
   {{m}\over {2\, K\, \sigma}}\,(K-1),\nonumber \\
   &&{}\nonumber \\
 \mbox{ or }\qquad&&| \Omega^r(\tau ,\sigma ) | < {{m}\over { K\,
   \sigma}}\,(K-1).
 \label{IV14}
 \eea
\bigskip

This means that, while the linear velocities ${\dot x}^{\mu}(\tau
)$ and the translational accelerations ${\ddot x}^{\mu}(\tau )$
are arbitrary, the allowed rotations $R(\alpha ,\beta ,\gamma )$
on the leaves of the foliation {\it have the rotational
frequencies}, namely the angular velocities $\Omega^r(\tau ,\sigma
)$, {\it limited by an upper bound proportional to the minimum of
the linear velocity $v_l(\tau ) = {\dot x}_{\mu}(\tau )\, l^{\mu}$
orthogonal to the parallel hyper-planes}.

\bigskip

Instead of checking the conditions (\ref{I1}) on $g_{rs}(\tau
,\vec \sigma )$, let us write

\bea
 z^{\mu}(\tau ,\vec \sigma ) &=& \xi_l(\tau ,\vec
 \sigma )\, l^{\mu} - \sum_r\, \xi_r(\tau ,\vec \sigma )\,
 \epsilon^{\mu}_r,\nonumber \\
 &&{}\nonumber \\
 \xi_l(\tau ,\vec \sigma ) &=& \sgn\, z_{\mu}(\tau ,\vec \sigma )\,
 l^{\mu} = \sgn\, x_{\mu}(\tau )\, l^{\mu} = x_l(\tau ),\nonumber \\
&&\nonumber\\
 \xi_r(\tau ,\vec \sigma ) &=& \sgn\, z_{\mu}(\tau ,\vec \sigma )\,
 \epsilon^{\mu}_r = \sgn\, x_{\mu}(\tau )\, \epsilon^{\mu}_r +
 R^r{}_s(\tau ,\sigma )\, \sigma^s = x_{\epsilon\, r}(\tau ) + R^r{}_s(\tau
 ,\sigma )\, \sigma^s,
\label{IV15} \eea

\noindent so that we get

 \bea
 \partial_{\tau}\, \xi_l(\tau ,\vec \sigma ) &=& {\dot x}_l(\tau )
 = v_l(\tau ),\qquad \partial_r\, \xi_l(\tau ,\vec \sigma ) = 0,\nonumber \\
 &&{}\nonumber \\
 \partial_u\, \xi_r(\tau ,\vec \sigma ) &=& R^r{}_u(\tau ,\sigma )
 + \partial_uR^r{}_s(\tau ,\sigma )\, \sigma^s =\nonumber\\
&&\nonumber\\
&=& R^r{}_v(\tau ,\sigma
 )\, \Big[ \delta^v_u + \omega^v_{(u) w}(\tau ,\sigma )\,
 \sigma^w\Big] =\nonumber \\
\nonumber\\
&=& R^r{}_v(\tau ,\sigma )\, \Big[ \delta^v_u +
 \Phi_{uv}(\tau ,\sigma )\, {{\sigma^u\, \sigma^w}\over
 {\sigma}}\Big] \, {\buildrel {def}\over =}\, \Big(R(\tau
 ,\sigma )\, M(\tau ,\vec \sigma )\Big)_{ru},
 \label{IV16}
 \eea
\medskip

\noindent and let us show that $\sigma^A = (\tau ,\vec \sigma )
\mapsto (\xi_l(\tau ,\vec \sigma ), \xi_r(\tau ,\vec \sigma )\, )$
is a coordinate transformation with positive Jacobian \cite{19a}.
This will ensure that these foliations with parallel hyper-planes
are defined by embeddings such that $\sigma^A \mapsto x^{\mu} =
z^{\mu}(\tau ,\vec \sigma )$ is the inverse of an admissible
4-coordinate transformation $x^{\mu} \mapsto \sigma^A$.
\medskip

Therefore we have to study the Jacobian

\bea
 J(\tau ,\vec \sigma ) &=& \left( \begin{array}{cc} {{\partial\,
 \xi_l(\tau ,\vec \sigma )}\over {\partial \tau}} & {{\partial\,
 \xi_s(\tau ,\vec \sigma )}\over {\partial \tau}}\\
 {{\partial\, \xi_l(\tau ,\vec \sigma )}\over {\partial \sigma^r}}
 & {{\partial\, \xi_s(\tau ,\vec \sigma )}\over {\partial \sigma^r}}
\end{array} \right) = \left( \begin{array}{cc} v_l(\tau )& {{\partial\,
 \xi_s(\tau ,\vec \sigma )}\over {\partial \tau}}\\
 0_r&  \Big(R(\tau ,\sigma )\, M(\tau ,\vec \sigma )\Big)_{rs}
\end{array} \right),\nonumber \\
 &&{}\nonumber \\
  &&{}\nonumber \\
   \det\, J(\tau ,\vec \sigma ) &=&
v_l(\tau )\, \det\, R(\tau ,\sigma )\, \det\, M(\tau ,\vec \sigma
) = v_l(\tau )\, \det\, M(\tau ,\vec \sigma ).
 \label{IV17}
 \eea
\medskip

To show that $\det\, M(\tau ,\vec \sigma ) \not= 0$, let us look
for the null eigenvectors $W_r(\tau ,\vec \sigma )$ of the matrix
$M(\tau ,\vec \sigma )$, $M_{rs}(\tau ,\vec \sigma )\, W_s(\tau
,\vec \sigma ) = 0$ or $W_r(\tau ,\vec \sigma ) - \Phi_{uv}(\tau
,\sigma ) \, {{\sigma^u}\over {\sigma}}\, \sigma^s\, W_s(\tau
,\vec \sigma ) = 0$ [see Eq.(\ref{IV3})]. Due to $\Phi_{uv} = -
\Phi_{vu}$, we get $\sigma^s\, W_s(\tau ,\vec \sigma ) = 0$ and
this implies $W_r(\tau ,\vec \sigma ) = 0$, i.e. the absence of
null eigenvalues. Therefore $\det\, M(\tau ,\vec \sigma ) \not= 0$
and an explicit calculation shows that $\det\, M(\tau ,\vec \sigma
) = 1$. As a consequence, we get $\det\, J(\tau ,\vec \sigma ) =
v_l(\tau ) > 0$. Therefore, $x^{\mu} \mapsto \sigma^A$ is an
admissible 4-coordinate transformation.

\bigskip

Let us remark that the congruence of time-like world-lines
associated to the constant normal $l^{\mu}$ defines an inertial
reference frame: each inertial observer is naturally endowed with
the orthonormal tetrad $b^{\mu}_A = (l^{\mu}; \epsilon^{\mu}_r)$.

\bigskip

Let us consider the second skew congruence, whose observer
world-lines are $x^{\mu}_{\vec \sigma}(\tau ) = z^{\mu}(\tau ,\vec
\sigma )$, and let us look for an orthonormal tetrad
$V^{\mu}_A(\tau ,\vec \sigma ) = (z^{\mu}_{\tau}(\tau ,\vec \sigma
)/\sqrt{\sgn\, g_{\tau\tau}(\tau ,\vec \sigma)}; V^{\mu}_r(\tau
,\vec \sigma ))$ to be associated to each of its time-like
observers. Due to the orthonormality we have $V^{\mu}_A(\tau ,\vec
\sigma ) = \Lambda^{\mu}{}_{\nu = A}(\tau ,\vec \sigma )$ with
$\Lambda (\tau ,\vec \sigma )$ a Lorentz matrix. Therefore we can
identify them with $SO(3,1)$ matrices parametrized as the product
of a pure boost with a pure rotation as in Eqs. (\ref{III8}). If
we introduce

\begin{eqnarray}
&&\vec{E}_r(\tau,\vec{\sigma})=\{E^k_r(\tau,\vec{\sigma})\}=R^{s=k}_r(\alpha_m
(\tau ,\sigma ),\beta_m (\tau ,\sigma ),\gamma_m (\tau ,\sigma ))
\nonumber\\
 &&{}\nonumber \\
 && \Rightarrow\,\, {{\partial {\vec E}_r(\tau ,\vec \sigma
 )}\over {\partial \tau}} = \stackrel{def}{=}
 \vec{\omega}_{m}
 (\tau) \times\vec{E}_r(\tau,\vec{\sigma}),\nonumber \\
 &&\nonumber\\
 &&B^{jk}(\vec{\beta}_m(\tau,\vec{\sigma}))=\delta^{ij}+
\frac{\gamma^2(\vec{\beta}_m(\tau ,\sigma
))}{\gamma(\vec{\beta}_m(\tau ,\sigma ))+1}\,\beta^i_m(\tau
,\sigma )\, \beta^j_m(\tau ,\sigma ),
 \label{IV18}
\end{eqnarray}

\noindent  we can write

\begin{equation}
V^\mu_{A}(\tau,\vec{\sigma})=\Lambda^\mu_{\nu=A}
(\tau,\vec{\sigma})= \left(
\begin{array}{cc}
\frac{1}{\sqrt{1-\vec{\beta}_m^2(\tau,\vec{\sigma})}}&
\frac{\vec{\beta}_m(\tau,\vec{\sigma})\cdot\vec{E}_{
r}(\tau,\vec{\sigma})}{\sqrt{1-\vec{\beta}_m^2(\tau,\vec{\sigma})}}\\
\frac{\beta^j_m(\tau,\vec{\sigma})}{\sqrt{1-\vec{\beta}_m^2(\tau,\vec{\sigma})}}&\,
B^{jk}({\vec \beta}_m (\tau,\vec{\sigma}))\,
E^k_{r}(\tau,\vec{\sigma})
\end{array}
\right).
 \label{IV19}
\end{equation}

We stress that for every observer $x^\mu_{\vec{\sigma}}(\tau)$ the
choice of the $V^\mu_r(\tau,\vec{\sigma})$'s, and therefore also
of the $\vec{E}_r(\tau,\vec{\sigma})$'s, is arbitrary. As a
consequence the angular velocity $\vec{\omega}_m(\tau)$ {\em
defined} by the second of the Eqs.(\ref{IV18}) is in general not
related with the angular velocity (\ref{IV2}) defined by the
embedding. On the contrary, the parameter
$\vec{\beta}_m(\tau,\vec{\sigma})$ is related to the embedding by
the relation $\beta^i_m(\tau,\vec{\sigma}) =
z^i_\tau(\tau,\vec{\sigma}) /z^o_\tau(\tau,\vec{\sigma})$.

\bigskip

For every observer $x^{\mu}_{\vec \sigma}(\tau )$ of the
congruence, endowed with the orthonormal tetrad $E^{\mu}_{\vec
\sigma\, A}(\tau ) = V^{\mu}_A(\tau ,\vec \sigma )$, we get

\bea
 &&
 {{d E^{\mu}_{\vec \sigma\, A}(\tau )}\over {d\tau}} =
 {{\cal A}_{\vec \sigma\, A}}^B(\tau )\, V^{\mu}_{\vec \sigma\, B}(\tau ),
 \nonumber \\
 &&{}\nonumber \\
 &\Rightarrow & {\cal A}_{\vec \sigma \, AB}(\tau ) = - {\cal
 A}_{\vec \sigma\, BA}(\tau ) = {{d E^{\mu}_{\vec \sigma\, A}(\tau
 )}\over {d\tau}}\,\eta_{\mu\nu}
 E^\nu_{\vec \sigma\, B}(\tau ),
 \label{IV20}
  \eea

Using the (\ref{IV19}) we obtain [$\gamma(\tau,\vec{\sigma}) = 1/
\sqrt{1 - {\vec \beta}_m^2(\tau,\vec{\sigma})}$,
$\dot{\vec{\beta}}_m(\tau,\vec{\sigma})=
d{\vec{\beta}}_m(\tau,\vec{\sigma})/d\tau$]

 \bea
 a_{\vec \sigma\, r}(\tau ) &=& {\cal A}_{\vec \sigma\, \tau
 r}(\tau ) =\left[
 -\gamma \,(\dot{\vec{\beta}}_m\cdot\vec{E}_r)-
 \frac{\gamma^3}{\gamma+1}\,
 (\dot{\vec{\beta}}_m\cdot\vec{\beta}_m)
 (\vec{\beta}_m\cdot\vec{E}_r)\right](\tau,\vec{\sigma})
 \nonumber \\
 &&\nonumber\\
 \Omega_{\vec \sigma\, r}(\tau ) &=& {1\over 2}\, \epsilon_{ruv}\,
 {\cal A}_{\vec \sigma\, uv}(\tau ) =\nonumber\\
 &&\nonumber\\
 &=&
 \left[
 -\vec{\omega}_m\cdot\vec{E}_r-
 \frac{\gamma^2}{\gamma+1}
 \epsilon^{rsu}(\vec{\beta}_m\cdot\vec{E}_s)
 (\dot{\vec{\beta}}_m\cdot\vec{E}_u)
 \right](\tau,\vec{\sigma})
 \label{IV21}
 \eea

Therefore the acceleration radii (see the Introduction) of these
observers are

\bea
 I_1 &=&{\vec \Omega}_{\vec \sigma}^2 - {\vec a}_{\vec \sigma}^2 =
 \left[
 \vec{\omega}^2_m+
 2\frac{\gamma^2}{\gamma+1}
 \vec{\omega}_m\cdot(\dot{\vec{\beta}}_m\times\vec{\beta}_m)+
 \gamma^2(\gamma-2)\,\dot{\vec{\beta}}_m^2-
 \frac{\gamma^6}{\gamma+1}\,
 (\dot{\vec{\beta}}_m\cdot\vec{\beta}_m)^2
 \right](\tau,\vec{\sigma})
 \nonumber \\
 &&{}\nonumber \\
 I_2 &=& {\vec a}_{\vec \sigma} \cdot {\vec \Omega}_{\vec \sigma} = \left[
 \gamma\,(\dot{\vec{\beta}}_m\cdot\vec{\omega}_m)+
 \frac{\gamma^3}{\gamma+1}\,
 (\dot{\vec{\beta}}_m\cdot\vec{\beta}_m)
 (\vec{\beta}_m\cdot\vec{\omega}_m)
 \right](\tau,\vec{\sigma})
 \label{IV22}
 \eea
\medskip

Let us remark that, even if we have finite acceleration radii, our
radar 4-coordinates are globally defined, differently from Fermi
coordinates.

 \bigskip

The non-relativistic limit of the embedding (\ref{IV1}) can be
obtained by choosing $\epsilon^{\mu}_r = (0; e^i_r)$. We obtain a
generalization of  the standard translating and rotating
3-coordinate systems on the hyper-planes of constant absolute
Newtonian time

\bea
 t^{'}(\tau ) &=& t(\tau ),\nonumber \\
 z^i(\tau ,\vec \sigma ) &=& x^i(\tau ) + e^i_r\, R^r{}_s(\tau
 ,\sigma )\, \sigma^s,
 \label{IV23}
 \eea

\noindent without any restriction on rotations, namely with $R =
R(\tau )$ allowed.

\vfill\eject

\section{The Modification of Einstein ${1\over 2}$ Convention
Associated to Admissible Radar Coordinates and their Empirical
Determination.}

In the previous Section we have seen how it possible to build as
many conventional observer-dependent globally defined radar
4-coordinate systems as admissible 3+1 splittings of space-time.
Many of them have the {\it equal-time} leaves, describing the
conventional instantaneous present, not orthogonal to the
world-line of the accelerated observer.
\medskip

In this Section we show that each admissible 3+1 splitting of
space-time leads to a different generalization (see also Havas
\cite{30}) of Einstein ${1\over 2}$ convention for the
synchronization of distant clocks by means of light signals.
\medskip

Then we show how such generalizations can lead to an operational
method for an empirical determination of admissible radar
4-coordinates around the world-line of a non-inertial observer
simulated by a spacecraft belonging to a cluster of spacecrafts
(or satellites) like the one used in the Global Positioning
System.

\subsection{A Cluster of Spacecrafts like in the Global Positioning System. }

In Eqs.(\ref{IV1}) we gave a family of embeddings $x^{\mu} =
z^{\mu}(\tau ,\vec \sigma )$ defining possible notions of
simultaneity, i.e. admissible 3+1 splittings of Minkowski
space-time with foliations with space-like hyper-planes
$\Sigma_{\tau}$ as leaves, to be associated to the world-line
$x^{\mu}(\tau )$ of an arbitrary time-like observer $\gamma$,
chosen as origin of the 3-coordinates on each simultaneity leaf
$\Sigma_{\tau}$, i.e. $x^{\mu}(\tau ) = z^{\mu}(\tau ,\vec 0)$
(with this definition in general $\tau$ is not the proper time of
the observer, but $(\tau ,\vec \sigma )$ are a good set of
observer-dependent radar coordinates). The space-like hyper-planes
$\Sigma_{\tau}$ are not orthogonal to $\gamma$: if $l^{\mu} =
\epsilon^{\mu}_{\tau}$ is the normal to $\Sigma_{\tau}$ we have
$l_{\mu}\, {{{\dot x}^{\mu}(\tau )}\over {\sqrt{\sgn\, {\dot
x}^2(\tau )}}} \not= \sgn$ except in the limiting case of an
inertial observer with 4-velocity proportional to $l^{\mu}$.
\medskip

If $\tau$ is not the proper time of the observer, the proper time
${\cal T}_{\gamma}$ of the standard atomic clock $C$ of $\gamma$
will be defined by $d{\cal T}_{\gamma} = \sqrt{\sgn\,
g_{\tau\tau}(\tau , \vec 0)}\, d\tau$ [$x^{\mu}(\tau ) = {\tilde
x}^{\mu}({\cal T}_{\gamma})$]. This defines ${\cal T}_{\gamma} =
{\cal F}_{\gamma}(\tau )$ as a monotonic function of $\tau$, whose
inverse will be denoted $\tau = {\cal G}({\cal T}_{\gamma})$.
Moreover, we make an arbitrary conventional choice of a tetrad
${}_{(\gamma )}E^{\mu}_A(\tau )$ associated to $\gamma$ with
${}_{(\gamma )}E^{\mu}_{\tau}(\tau ) = {{{\dot x}^{\mu}(\tau )
}\over {\sqrt{\sgn\, {\dot x}^2(\tau )}}}$.

\bigskip

Let us consider a set of $N$ arbitrary time-like world-lines
$x^{\mu}_i(\tau )$, $i=1,..,N$, associated to observers
$\gamma_i$, so that $\gamma$ and the $\gamma_i$'s can be imagined
to be the world-lines of $N + 1$ spacecrafts (like in GPS
\cite{18}) with $\gamma$ chosen as a reference world-line (the
accelerated observer). Each of the world-lines $\gamma_i$ will
have an associated standard atomic clock $C_i$ and a conventional
tetrad ${}_{(\gamma_i)}E^{\mu}_A(\tau )$.

\medskip

To compare the distant clocks $C_i$ with $C$ in the chosen notion
of simultaneity, we define the 3-coordinates ${\vec \eta}_i(\tau
)$ of the $\gamma_i$

\beq
 x^{\mu}_i(\tau ) {\buildrel {def}\over =} z^{\mu}(\tau ,{\vec \eta}_i(\tau )).
 \label{VI1}
 \eeq

\noindent Then the proper times ${\cal T}_{\gamma_i}$ of the
clocks $C_i$ will be expressed in terms of the scalar coordinate
time $\tau$ of the chosen simultaneity as

\beq
 d{\cal T}_{\gamma_i} = \sqrt{\sgn\, \Big[ g_{\tau\tau}(\tau
,{\vec \eta}_i(\tau ))+ 2g_{\tau r}(\tau ,{\vec \eta}_i(\tau
))\,\dot{\eta}^r(\tau)+ g_{rs}(\tau ,{\vec \eta}_i(\tau
))\,\dot{\eta}^r(\tau)\,\dot{\eta}^s(\tau) \Big]}\, d\tau,
 \label{VI2}
  \eeq

\noindent so that with this notion of simultaneity the proper
times ${\cal T}_{\gamma_i}$ are connected to the proper time
${\cal T}_{\gamma}$ by the following relations

\beq
 d{\cal T}_{\gamma_i} = \left.\sqrt{{{g_{\tau\tau}(\tau ,{\vec \eta}_i(\tau ))+
2g_{\tau r}(\tau ,{\vec \eta}_i(\tau ))\,\dot{\eta}_i^r(\tau)+
g_{rs}(\tau ,{\vec \eta}_i(\tau
))\,\dot{\eta}_i^r(\tau)\,\dot{\eta}_i^s(\tau) } \over
{g_{\tau\tau}(\tau ,\vec 0)}}}\, \right|_{\tau = {\cal G}({\cal
 T}_{\gamma})}\, d{\cal T}_{\gamma}.
 \label{VI3}
 \eeq
\medskip

This determines the synchronization of the $N + 1$ clocks once we
have expressed the 3-coordinates ${\vec \eta}_i(\tau )$ in terms
of the given world-lines $x^{\mu}(\tau )$, $x^{\mu}_i(\tau )$ and
of an admissible embedding. For the embedding (\ref{IV1}) from the
definition

\beq
 x_i^{\mu}(\tau ) =
 z^{\mu}(\tau ,{\vec \eta}_i(\tau )) = x^{\mu}(\tau ) +
  \epsilon^{\mu}_r\, R^r{}_s(\tau ,|{\vec \eta}_i(\tau )|)\, \eta^s_i(\tau ),
 \label{VI4}
 \eeq
\medskip

\noindent   we get [$\,\,\, |{\vec \eta}_i(\tau )| \, {\buildrel
{def}\over =}\, \sqrt{\delta_{rs}\, \eta^r_i(\tau )\,
\eta^s_i(\tau )}$, $\eta^r_i(\tau ) =  |{\vec \eta}_i(\tau )|\,
{\hat  n}_i^r(\tau )$, $\delta_{rs}\, {\hat n}^r_i(\tau )\, {\hat
n}^s_i(\tau ) = 1$]
\medskip

\beq
  \eta^u_i(\tau ) = - \sum_w\, [R^{-1}(\tau , |{\vec \eta}_i(\tau
  )|)]^u{}_w\, \epsilon^{\nu}_w\, [x_{i\nu}(\tau ) - x_{\nu}(\tau
  )].
 \label{VI5}
 \eeq

Then, if we put the solution

\beq
 |{\vec \eta}_i(\tau )| =
  F_i\Big[\epsilon^{\mu}_r\, \Big(x_{i\mu}(\tau ) - x_{\mu}(\tau
  )\Big)\Big],
 \label{VI6}
 \eeq

\noindent of the equations

\bea
 |{\vec \eta}_i(\tau )|^2 &=& \delta_{rs}\,
 \sum_{mn}\,  [R^{-1}(\tau ,|{\vec \eta}_i(\tau
  )|)]^r{}_m\, [R^{-1}(\tau ,|{\vec \eta}_i(\tau
  )|)]^s{}_n\nonumber \\
  &&\epsilon^{\mu}_m\, [x_{i\mu}(\tau ) - x_{\mu}(\tau )]\,
  \epsilon^{\nu}_n\, [x_{i\nu}(\tau ) - x_{\nu}(\tau )],
 \label{VI7}
 \eea

\noindent into Eqs.(\ref{VI5}), we obtain the looked for
expression of the 3-coordinates ${\vec \eta}_i(\tau )$

  \beq
\eta^u_i(\tau ) = - \sum_m\, \Big[ R^{-1}(\tau ,
  F_i[\epsilon^{\alpha}_w\, (x_{i\alpha}(\tau ) - x_{\alpha}(\tau
  ))])\Big]^u{}_m\, \epsilon^{\nu}_m\, [x_{i\nu}(\tau ) -
  x_{\nu}(\tau )].
 \label{VI8}
 \eeq

\subsection{The Modification of Einstein ${1\over 2}$ Convention
for a M$\o$ller Admissible Radar Coordinate System.}

Let us now consider an admissible embedding $x^{\mu} = z^{\mu}
(\tau ,\vec \sigma )$ of the family (\ref{IV1}) (but the
discussion applies to every admissible embedding), where $(\tau
,\vec \sigma )$ are the radar coordinates adapted to the
accelerated observer with world-line $\gamma$. On the simultaneity
leave $\Sigma_{\tau}$ having the point $Q$ of 4-coordinates
$x^{\mu}(\tau )$ on $\gamma$ as origin of the 3-coordinates $\vec
\sigma$, let us consider a point $P$ on $\Sigma_{\tau}$ with
coordinates $z^{\mu}(\tau ,\vec \sigma )$ (for $\vec \sigma =
{\vec \eta}_i(\tau )$ it corresponds to the spacecraft
$\gamma_i$). {\it We want to express the observer-dependent radar
4-coordinates $\tau = \tau (P)$, $\vec \sigma = \vec \sigma (P)$
of $P$ in terms of data on the world-line $\gamma$ corresponding
to the emission of a light signal in $Q_{-}$ at $\tau_{-} < \tau$
and to its reception in $Q_{+}$ at $\tau_{+} > \tau$ after
reflection at $P$}.
\medskip

Let $x^{\mu}(\tau_{-})$ be the intersection of the world-line
$\gamma$ with the past light-cone through $P$ and
$x^{\mu}(\tau_{+})$ the intersection with the future light-cone
through $P$. To find $\tau_{\pm}$ we have to solve the equations
$\Delta^2_{\pm} = [x^{\mu}(\tau_{\pm}) - z^{\mu}(\tau ,\vec \sigma
)]^2 = 0$ with $\Delta^{\mu}_{\pm} = x^{\mu}(\tau_{\pm}) -
z^{\mu}(\tau ,\vec \sigma )$. We are interested in the solutions
$\Delta^o_{+} = |{\vec \Delta}_{+}|$ and $\Delta^o_{-} = - |{\vec
\Delta}_{-}|$. Let us remark that on the simultaneity surfaces
$\Sigma_{\tau}$ we have $x^o(\tau ) \not= z^o(\tau ,\vec \sigma )$
for the Cartesian coordinate times.

\bigskip

Let us show that the $\gamma$-dependent radar coordinates $\tau$
and $\vec \sigma$ of the event $P$, with Cartesian 4-coordinates
$z^{\mu}(\tau ,\vec \sigma )$ in an inertial system, can be
determined in terms of the {\it emission scalar time} $\tau_{-}$
of the light signal, the {\it emission unit 3-direction} ${\hat
n}_{(\tau_{-})}( \theta_{(\tau_{-})} ,\phi_{(\tau_{-})} )$ [so
that $\triangle^{\mu}_{-} = |{\vec \triangle}_{-}|\, (-\sgn ;
{\hat n}_{(\tau_{-})})$ ] and the {\it reception scalar time}
$\tau_{+}$ registered by the observer $\gamma$ with world-line
$x^{\mu}(\tau )$ \cite{20a}. These data are usually given in terms
of the proper time ${\cal T}(\tau )$ of the observer $\gamma$ by
using $d{\cal T} = \sqrt{\sgn\, g_{\tau\tau}(\tau ,\vec 0) }\,
d\tau$.
\bigskip

Let us introduce the following parametrization by using
Eqs.(\ref{IV1})

\begin{eqnarray*}
 z^{\mu}(\tau ,\vec \sigma ) &=& x^{\mu}(\tau ) +
 \epsilon^{\mu}_r\, R^r{}_s(\tau ,\sigma )\, \sigma^s\, {\buildrel
 {def}\over =}\nonumber \\
 &&{}\nonumber \\
 &{\buildrel {def}\over =}& \Big[ \xi_l(\tau ,\vec
 \sigma )\, l^{\mu} + \xi^r(\tau ,\vec \sigma )\,
 \epsilon^{\mu}_r\Big] =\nonumber \\
&&\nonumber\\
 &=&  \Big[ x_l(\tau )\, l^{\mu} + \sum_r\, [x_{\epsilon}^r(\tau ) + \zeta^r(\tau
 ,\vec \sigma )]\, \epsilon^{\mu}_r\Big],\nonumber \\
 &&{}\nonumber \\
 \xi_l(\tau ,\vec \sigma ) &=& \sgn\,z_{\mu}(\tau ,\vec \sigma )\, l^{\mu} =
 \sgn\, x_{\mu}(\tau )\, l^{\mu} = x_l(\tau ),
 \end{eqnarray*}

\bea
  \xi^r(\tau ,\vec \sigma ) &=& \sgn\,z_{\mu}(\tau ,\vec \sigma )\,
  \epsilon^{\mu}_r = x^r_{\epsilon}(\tau ) + \zeta^r(\tau ,\vec
  \sigma ),\nonumber \\
&&\nonumber\\
  x_{\epsilon}^r(\tau ) &=&\sgn\,
 x_{\mu}(\tau )\, \epsilon^{\mu}_r,\qquad \zeta^r(\tau ,\vec
 \sigma ) = R^r{}_s(\tau ,\sigma )\, \sigma^s\,
 {\rightarrow}_{\sigma \rightarrow \infty}\, \sigma^r.
 \label{VI9}
 \eea

\medskip

Then the two equations $\triangle^2_{\pm} = [x^{\mu}(\tau_{\pm}) -
z^{\mu}(\tau ,\vec \sigma )]^2 = \sgn\, ([x_l(\tau_{\pm}) -
x_l(\tau )]^2 - [{\vec x}_{\epsilon}(\tau_{\pm}) - {\vec
x}_{\epsilon}(\tau ) - \vec \zeta (\tau ,\vec \sigma )]^2) = 0$
can be rewritten in the form

 \bea
  x_l(\tau_{+}) &=& x_l(\tau ) + |{\vec \triangle}_{+}|
  = x_l(\tau ) + | {\vec x}_{\epsilon}(\tau_{+}) - \vec \xi
  (\tau ,\vec \sigma )|,\nonumber \\
   &&\qquad |{\vec \Delta}_{+}| = |{\vec x}_{\epsilon}(\tau_{+}) -
  \vec \xi (\tau ,\vec \sigma )|\, {\rightarrow}_{\sigma \rightarrow \infty}\,
  |{\vec x}_{\epsilon}(\tau_{+}) - {\vec x}_{\epsilon}(\tau ) - \vec \sigma |,\nonumber \\
  &&{}\nonumber \\
  x_l(\tau_{-}) &=& x_l(\tau ) - |{\vec \triangle}_{-}|
  = x_l(\tau ) - | {\vec x}_{\epsilon}(\tau_{-}) - \vec \xi
  (\tau ,\vec \sigma )|,\nonumber \\
   &&\qquad |{\vec \Delta}_{-}| = |{\vec x}_{\epsilon}(\tau_{-}) -
  \vec \xi (\tau ,\vec \sigma )|\, {\rightarrow}_{\sigma \rightarrow \infty}\,
  |{\vec x}_{\epsilon}(\tau_{-}) - {\vec x}_{\epsilon}(\tau ) - \vec \sigma |.
  \label{VI10}
  \eea

It can be shown \cite{42} that, if no observer is allowed to
become a Rindler observer \cite{41}, then each equation admits a
unique \cite{21a} solution $\tau_{\pm} = T_{\pm}(\tau ,\vec \sigma
)$.

Therefore the following four data measured by the observer
$\gamma$

  \begin{eqnarray*}
  \tau_{\pm} &=& T_{\pm}(\tau ,\vec \sigma ),\nonumber \\
&&{}\nonumber \\
 \end{eqnarray*}

\bea
  {\hat n}_{(\tau_{-})}(\theta_{(\tau_{-})}, \phi_{(\tau_{-})})
  &=& \Big( \sin\, \theta_{(\tau_{-})}\, \sin\, \phi_{
  (\tau_{-})}, \sin\, \theta_{(\tau_{-})}\, \cos\, \phi_{(\tau_{-})},
  \cos\, \theta_{(\tau_{-})}\Big) =\nonumber \\
&&\nonumber\\
  &=& {{{\vec \triangle}_{-}}\over {|{\vec \triangle}_{-}|}}
  = \left.{{{\vec x}_{\epsilon}(\tau_{-}) - {\vec x}_{\epsilon}(\tau )
  - \vec \zeta (\tau ,\vec  \sigma )}\over {|{\vec x}
  _{\epsilon}(\tau_{-}) - {\vec x}_{\epsilon}(\tau ) - \vec \zeta (\tau ,\vec
  \sigma )|}}\right|_{\tau_{-} = T_{-}(\tau ,\vec \sigma )} = \hat m(\tau ,\vec \sigma ),
  \label{VI11}
  \eea

\noindent can be inverted to get the adapted coordinates $\tau
(P)$, $\vec \sigma (P)$ of the event $P$ with 4-coordinates
$z^{\mu}(\tau ,\vec \sigma )$ in terms of the data (Einstein's
convention for the radar time would be ${\cal E} = {1\over 2}$)

\bea
 \tau (P) &=& \tau (\tau_{-}, {\hat n}_{(\tau_{-})}, \tau_{+})
{\buildrel {def}\over =} \tau_{-} + {\cal E}(\tau_{-}, {\hat
n}_{(\tau_{-})}, \tau_{+})\, [\tau_{+} - \tau_{-}],\nonumber \\
 &&{}\nonumber \\
 \vec \sigma (P)&=& {\vec {\cal G}}(\tau_{-}, {\hat n}_{(\tau_{-})},
 \tau_{+})\, {\rightarrow}_{\tau_{+} \rightarrow \tau_{-}}\, 0.
 \label{VI12}
 \eea

\bigskip

Let us remark that

i) for $x^{\mu}(\tau ) = \tau\, l^{\mu}$ (inertial observer with
world-line {\it orthogonal} to $\Sigma_{\tau}$; ${\vec
x}_{\epsilon}(\tau ) = 0$) we get the Einstein's convention for
the radar time, because we have
\medskip

\begin{eqnarray*}
 \tau_{\pm} &=& \tau \pm |\vec \zeta (\tau ,\vec \sigma )|,\qquad \tau =
\tau_{-} + {1\over 2}\, (\tau_+ - \tau_{-}) = {1\over 2}\,
(\tau_{+} + \tau_{-}),\qquad  {\cal E} = {1\over 2},\nonumber \\
 \sigma &=& |\vec \zeta (\tau ,\vec \sigma )| = {1\over 2}\,
(\tau_{+} - \tau_{-}), \qquad \zeta^r(\tau ,\vec \sigma ) = -
{1\over 2}\, (\tau_{+} - \tau_{-})\, {\hat
n}^r_{(\tau_{-})},\nonumber \\
 \sigma^r &=& {\cal G}^r = {1\over
2}\, (\tau_{+} - \tau_{-})\, (R^{-1})^r{}_s({{\tau_{+} +
\tau_{-}}\over 2}, {{\tau_{+} - \tau_{-}}\over 2})\, {\hat
n}^s_{(\tau_{-})};
 \end{eqnarray*}

ii) for $x^{\mu}(\tau ) = \tau\, [l^{\mu} + \epsilon^{\mu}_r\,
a^r]$ (inertial observer with world-line {\it non-orthogonal} to
$\Sigma_{\tau}$; ${\vec x}_{\epsilon}(\tau ) = \tau\, \vec a$),
after some straightforward calculations, we get

\begin{eqnarray*}
 \tau_{\pm} &=& \tau + {1\over {1 - {\vec a}^2}}\, \Big[- \vec a \cdot
\vec \zeta (\tau ,\vec \sigma ) \pm \sqrt{(\vec a \cdot \vec \zeta
 (\tau ,\vec \sigma ) )^2 + (1 - {\vec a}^2)\, \sigma^2}\Big],\nonumber \\
 \tau &=& {1\over 2}\, \Big[\tau_{+} + \tau_{-} + {{\tau_{+} -
\tau_{-}}\over {1 - {\vec a}^2}}\, \sqrt{{{{\vec a}^2 + \vec a
\cdot {\hat n}_{(\tau_{-})}}\over {1 + {\vec a}^2 - {\vec a}^4 +
(3 - 2\, {\vec a}^2)\, \vec a \cdot {\hat n}_{(\tau_{-}})
}}}\Big],\nonumber \\
 {\cal E} &=& {1\over 2}\, \Big[ 1 + {1\over {1 - {\vec a}^2}}\,
\sqrt{{{{\vec a}^2 + \vec a \cdot {\hat n}_{(\tau_{-})}}\over {1 +
{\vec a}^2 - {\vec a}^4 + (3 - 2\, {\vec a}^2)\, \vec a \cdot
{\hat n}_{(\tau_{-}}) }}}\Big],
 \end{eqnarray*}

\begin{eqnarray*}
 \sigma &=& |\vec \zeta (\tau ,\vec \sigma )| = {1\over 2}\,
(\tau_{+} - \tau_{-})\,  \sqrt{{{1 + {\vec a}^2 + 2\, \vec a \cdot
{\hat n}_{(\tau_{-})}}\over {1 + {\vec a}^2 - {\vec a}^4 + (3 -
 2\, {\vec a}^2)\, \vec a \cdot {\hat n}_{(\tau_{-}}) }}},\nonumber \\
 {{\zeta^r(\tau ,\vec \sigma )}\over {|\vec \zeta (\tau ,\vec
\sigma )|}} &=& - {{  \sqrt{{\vec a}^2 + \vec a \cdot {\hat
n}_{(\tau_{-})}} + \sqrt{1 + {\vec a}^2 - {\vec a}^4 + (3 - 2\,
{\vec a}^2)\, \vec a \cdot {\hat n}_{(\tau_{-}}) } }\over {\sqrt{1
+ {\vec a}^2 + 2\, \vec a \cdot {\hat n}_{(\tau_{-})} }}}\, {{a^r
+ {\hat n}^r_{(\tau_{-})} }\over {1 - {\vec a}^2}},\nonumber \\
 \sigma^r &=& {\cal G}^r = - {1\over 2}\, (\tau_{+} - \tau_{-})\, \Big(
1 + \sqrt{ {{{\vec a}^2 + \vec a \cdot {\hat n}_{(\tau_{-})}}\over
{1 + {\vec a}^2 - {\vec a}^4 + (3 - 2\, {\vec a}^2)\, \vec a \cdot
{\hat n}_{(\tau_{-}}) }} }\Big)\nonumber \\
 &&(R^{-1})^r{}_s\Big({1\over 2}\,
\Big[\tau_{+} + \tau_{-} + {{\tau_{+} - \tau_{-}}\over {1 - {\vec
a}^2}}\, \sqrt{{{{\vec a}^2 + \vec a \cdot {\hat
n}_{(\tau_{-})}}\over {1 + {\vec a}^2 - {\vec a}^4 + (3 - 2\,
{\vec a}^2)\, \vec a \cdot {\hat n}_{(\tau_{-}})
}}}\Big],\nonumber \\
 &&{1\over 2}\, (\tau_{+} - \tau_{-})\,  \sqrt{{{1 + {\vec a}^2 + 2\,
\vec a \cdot {\hat n}_{(\tau_{-})}}\over {1 + {\vec a}^2 - {\vec
a}^4 + (3 - 2\, {\vec a}^2)\, \vec a \cdot {\hat n}_{(\tau_{-}})
}}} \Big)\quad {{a^s + {\hat n}^s_{(\tau_{-})}}\over {1 - {\vec
a}^2}};
\end{eqnarray*}

iii) for non-inertial trajectories $x^{\mu}(\tau ) = f(\tau )\,
l^{\mu} + \epsilon^{\mu}_r\, g^r(\tau )$ [$\sgn\, [{\dot f}^2(\tau
) - \sum_r\, {\dot g}^r(\tau )\, {\dot g}^r(\tau )] > 0$] the
evaluation of ${\cal E}$ and ${\vec {\cal G}}$ cannot be done
analytically, but only numerically.

\bigskip

In conclusion, each admissible 3+1 splitting of Minkowski
space-time together with an accelerated observer of world-line
$\gamma$ defines four functions ${\cal E}(\tau_{-}, {\hat
n}_{(\tau_{-})}, \tau_{+})$, ${\vec {\cal G}}(\tau_{-}, {\hat
n}_{(\tau_{-})}, \tau_{+})$ describing the modification of the
Einstein ${1\over 2}$ convention associated to the
observer-dependent radar coordinates describing the simultaneity
surfaces of the foliation. Since the admissible foliation of
Minkowski space-time is well defined at spatial infinity, the
functions ${\cal E}$ and ${\vec {\cal G}}$ will have a finite
limit for $\tau_{\pm}\, \rightarrow\, \pm \infty$, i.e. at spatial
infinity on $\Sigma_{\tau}$.

\bigskip

Let us add a remark on the {\it one-way} velocity of light
associated to these modifications of Einstein convention for the
synchronization of distant clocks.
\medskip

Given an admissible embedding $z^{\mu}(\tau ,\vec \sigma )$ and
its associated 4-metric $g_{AB}(\tau ,\vec \sigma )$, the {\em
instantaneous spatial distance} between two events $P$ and $Q$ on
the {\em same} hyper-surface $\Sigma_\tau$ is given by the {\em
line element on $\Sigma_{\tau}$}: $d\ell(\tau,\vec{\sigma})=
\sqrt{-\sgn\, g_{rs}(\tau,\vec{\sigma})\,d\sigma^r\,d\sigma^s}$.
If $P$ and $Q$ have a finite separation, their spatial distance is
obtained by integrating this line element along a geodesic on
$\Sigma_{\tau}$ connecting them, $\Gamma(P,Q)$: $D_\tau(P,Q) =
\int_{\Gamma(P,Q)}\,d\ell(\tau,\vec{\sigma})$.

If $Q$ is on the observer world-line $\gamma$, with coordinates
$(\tau ,\vec 0)$, and the nearby $P$ has coordinates $(\tau ,
d\vec \sigma )$, then we have $d\ell_\tau(P,Q)= \sqrt{-\sgn\,
g_{rs}(\tau,0)\,d\sigma^r\,d\sigma^s}$ ($g_{\tau\tau}(\tau,0)$
defines the relation between the $\tau$ and the observer proper
time  ${\cal T}_\gamma$). Then the coordinates of $P$ can be
parametrized in the form $d\sigma^r=\hat{n}^r\,d\ell_\tau(P,Q)$,
where $\hat{n}^r$ are the components  of a unit 3-vector [$\sgn\,
g_{rs}(\tau,0)\, \hat{n}^r\, \hat{n}^s = -1$].

\hfill

We have now all the elements to define the {\em instantaneous
velocity of light for a ray emitted by the observer $\gamma$ at
$Q$ (on $\Sigma_\tau$) in the direction $\hat{n}^r$}. If we assume
that the ray of light is received at the event $P$ with radar
coordinates
$\Big(\tau+d\tau_P,d\sigma^r=\hat{n}^r\,d\ell_\tau(P,Q)\Big)$, the
{\em instantaneous velocity of light} is $c_{-} =
\frac{d\ell_\tau(P,Q)}{d\tau_Q}$, if the infinitesimal {\it time
difference} is given by the future pointing solution of the
equation $g_{\tau\tau}(\tau,0)(d\tau_P)^2+ 2g_{\tau
r}(\tau,0)\,\hat{n}^r\,d\tau_P -[d\ell_\tau(P,Q)]^2 = 0$.
Therefore, we obtain the following {\em non-isotropic,
synchronization-dependent} velocity of light

\begin{equation}
c_{-}=g_{\tau r}(\tau,0)\,\hat{n}^r+\sqrt{ (g_{\tau
r}(\tau,0)\,\hat{n}^r)^2+g_{\tau\tau}(\tau,0) }.
 \label{AA}
 \end{equation}

\hfill

Following Synge \cite{43} and using the functions $T_{\pm}(\tau
,\vec \sigma )$ of Eqs.(\ref{VI11}), we can show that we have
$[d\ell_\tau(P,Q)]^2=\frac{1}{g_{\tau\tau}(\tau,0)}
\,\frac{\partial T^+}{\partial\sigma^r}(\tau,0) \,\frac{\partial
T^-}{\partial\sigma^s}(\tau,0) d\sigma^r\,d\sigma^s$  and
$g_{r\tau}(\tau,0)=\frac{1}{2g_{\tau\tau}(\tau,0)} \left(
\,\frac{\partial T^+}{\partial\sigma^r}(\tau,0)+ \,\frac{\partial
T^-}{\partial\sigma^s}(\tau,0) \right)$. As a consequence, an
observer choice of the synchronization convention, i.e. of
$T^{\pm}(\tau,\vec{\sigma})$, determines both the infinitesimal
spatial distance and the instantaneous velocity of light.

\hfill

Instead, in Section 4 of Ref.\cite{28}, there is an attempt to
define a {\it mean spatial distance and mean one-way light
velocities}, when the observer $\gamma$ emits a ray of light at
$\tau_-$ (event $Q_{-}$) and reabsorbs it at $\tau_+$ (event
$Q_+$) after a reflection at $P$. If we call $|{\vec \Delta}_{-}|$
the {\it light distance of $Q_{-}$ on $\gamma$ to $P$} and $|{\vec
\Delta}_{+}|$ the {\it light distance of $P$ to $Q_{+}$ on
$\gamma$ } (these are {\it mean} distances) we get the following
two {\it mean one-way velocities of light} (with $c = 1$) in
coordinates adapted to the given notion of simultaneity

\bea
 c_{-} &=& {{|{\vec \Delta}_{-}|}\over {\tau - \tau_{-}}} =
{{|{\vec \Delta}_{-}|}\over {{\cal E}\, (\tau_{+} - \tau_{-})}} =
{{2\, \eta\, |\vec \Delta | }\over {{\cal E}\, (\tau_{+} -
\tau_{-})}},
 \;\;\qquad from\, Q_{-}\, to\, P,\nonumber \\
 &&{}\nonumber \\
 c_{+} &=& {{|{\vec \Delta}_{+}|}\over {\tau_{+} - \tau}} =
{{|{\vec \Delta}_{+}|}\over {(1 - {\cal E})\, (\tau_{+} -
\tau_{-})}} = {{2\, (1 - \eta )\, |\vec \Delta | }\over {(1 -
{\cal E})\, (\tau_{+} - \tau_{-})}},\;\;\qquad from\, P\, to\, Q_{+},\nonumber \\
 &&{}\nonumber \\
 && |\vec \Delta|\, {\buildrel {def}\over =}\, {1\over 2}\,
  (|{\vec \Delta}_{+}| + |{\vec \Delta}_{-}|),\qquad \eta\,
  {\buildrel {def}\over =}\, {{|{\vec \Delta}_{-}|}\over
  {|{\vec \Delta}_{+}| + |{\vec \Delta}_{-}|}}.
  \label{VI16}
  \eea

\noindent If $c_{\tau} = {{2\, |\vec \Delta|}\over {\tau_{+} -
\tau_{-}}}$ is the isotropic average round-trip $\tau$-coordinate
velocity of light, we get $c_{+} = {{1 - \eta}\over {1- {\cal
E}}}\, c_{\tau}$, $c_{-} = {{\eta}\over {{\cal E}}}\, c_{\tau}$.
\medskip

If $x^{\mu}(\tau )$ is a straight-line (inertial observer) we can
adopt Einstein's convention ${\cal E} = {1\over 2}$, i.e. $\tau
(P) = {1\over 2}\, (\tau_{+} + \tau_{-})$ and $|\vec \sigma | =
|{\vec {\cal G}}| = {1\over 2}\, (\tau_{+} - \tau_{-})$
(hyper-planes orthogonal to the observer world-line). This implies
$|{\vec \Delta}_{+}| = |{\vec \Delta}_{-}|$ and $\eta = {1\over
2}$.

Instead, if we ask $c_{\tau} = c_{+} = c_{-}$, i.e. {\it isotropy}
of light propagation, we get ${\cal E} = \eta$. The conclusion of
Ref. \cite{28} is that once we have made a convention on {\it two}
of the quantities {\it spatial distance}, {\it one-way speed of
light} and {\it simultaneity}, the third one is automatically
determined. We have seen that at the level of exact, not mean,
quantities the convention about simultaneity and the associated
geodesic spatial distance on $\Sigma_{\tau}$ are enough to
determine the one-way velocity of light.

\subsection{The Inverse Problem and the Empirical Determination
of a Set of Radar Coordinates.}

We can now formulate an inverse problem: which are the
restrictions on four functions ${\cal E}(\tau_{-}, {\hat
n}_{(\tau_{-})}, \tau_{+})$, ${\vec {\cal G}}(\tau_{-}, {\hat
n}_{(\tau_{-})}, \tau_{+})$, so that they describe the
modification of Einstein ${1\over 2}$ convention for an
accelerated observer using the leaves of an admissible 3+1
splitting of Minkowski space-time  as simultaneity surfaces?

\bigskip

Let us  consider an infinitesimal displacement $\delta z^{\mu} =
z^{\mu}(\tau + \delta \tau , \vec \sigma + \delta \vec \sigma ) -
z^{\mu}(\tau ,\vec \sigma )$ of $P$ on $\Sigma_{\tau}$ to $P^{'}$
on $\Sigma_{\tau + \delta \tau}$. The event $P^{'}$ will receive
light signals from the event $Q(\tau_{-} + \delta \tau_{-})$ on
$\gamma$ and will reflect them towards the event $Q(\tau_{+} +
\delta \tau_{+})$ on $\gamma$. Now, using $\Delta_{\pm}^2 = 0$, we
have $\Delta^{'\, \mu}_{\pm} = \Delta^{\mu}_{\pm} + {\dot
x}^{\mu}(\tau_{\pm})\, \delta \tau_{\pm} - \delta z^{\mu}$ and
$\Delta^{'\, 2}_{\pm} = 2\, \Delta^{\mu}_{\pm}\, [{\dot
x}_{\mu}(\tau_{\pm})\, \delta \tau_{\pm} - \delta z_{\mu}] +
(higher\, order\, terms)$. As a consequence we get (see
Ref.\cite{9})

\bea
 {{\partial \tau_{\pm}}\over {\partial z^{\mu}}} &=& {{\Delta
 _{\pm\, \mu}}\over {\Delta_{\pm} \cdot {\dot x}(\tau_{\pm})}},\nonumber \\
 &&{}\nonumber \\
 with&& \sgn\, \Delta_{+} \cdot \Delta_{-} < 0,\quad
 \sgn\, \dot x(\tau_{+}) \cdot \Delta_{+} > 0,\quad
 \sgn\, \dot x(\tau_{-}) \cdot \Delta_{-} < 0.
 \label{VI13}
 \eea

\bigskip

Since ${{\partial \tau (P)}\over {\partial z^{\mu}}}$ is a
time-like 4-vector orthogonal to $\Sigma_{\tau}$, it must be
proportional to the normal $l^{\mu}$ to the space-like
hyper-planes of the foliation (\ref{IV1}) till now considered. For
a general admissible foliation we have (from $\triangle^2_{-} = 0$
we get $\triangle_{-} \cdot {{\partial \triangle_{-}}\over
{\partial z^{\mu}}} = 0$ and then $\triangle^{\mu}_{-}\,
{{\partial {\hat n}_{\tau_{-}}}\over {\partial z^{\mu}}} = 0$;
instead in general $\triangle^{\mu}_{+}\, {{\partial {\hat
n}_{\tau_{-}}}\over {\partial z^{\mu}}} \not= 0$)

\begin{eqnarray*}
 {{\partial \tau (P)}\over {\partial z^{\mu}}} &=& \Big[ {\cal E}
 + (\tau_{+} - \tau_{-})\, {{\partial {\cal E}}\over {\partial
 \tau_{+}}}\Big]\, {{\partial \tau_{+}}\over {\partial z^{\mu}}} +\nonumber \\
 &+& \Big[ 1 - {\cal E} + (\tau_{+} - \tau_{-})\,  {{\partial
 {\cal E}}\over {\partial \tau_{-}}}  \Big]\, {{\partial \tau_{-}}\over {\partial
 z^{\mu}}} +\nonumber \\
 &+& (\tau_{+} - \tau_{-})\, {{\partial {\cal E}}\over {\partial\,
 {\hat n}_{(\tau_{-})} }}\, {{\partial {\hat n}_{(\tau_{-})}}\over
 {\partial z^{\mu}}}  =\nonumber \\
 &=& \Big[ {\cal E}
 + (\tau_{+} - \tau_{-})\, {{\partial {\cal E}}\over {\partial
 \tau_{+}}}\Big]\,  {{\Delta
 _{+\, \mu}}\over {\Delta_{+} \cdot {\dot x}(\tau_{+})}}
 +\nonumber \\
 &+& \Big[ 1 - {\cal E} + (\tau_{+} - \tau_{-})\, \Big( {{\partial
 {\cal E}}\over {\partial \tau_{-}}} + {{\partial {\cal E}}\over {\partial\,
 {\hat n}_{(\tau_{-})}}}\, {{\partial {\hat n}_{(\tau_{-})}}\over
 {\partial \tau_{-}}}\Big) \Big]\,  {{\Delta
 _{-\, \mu}}\over {\Delta_{-} \cdot {\dot
 x}(\tau_{-})}} +\nonumber \\
 &+& (\tau_{+} - \tau_{-})\, {{\partial {\cal E}}\over {\partial\,
 {\hat n}_{(\tau_{-})} }}\, {{\partial {\hat n}_{(\tau_{-})}}\over
 {\partial z^{\mu}}} ,
 \end{eqnarray*}

 \bea
 \sgn\, \Big({{\partial \tau (P)}\over {\partial z^{\mu}}}\Big)^2
 &=& \sgn\, {{\Delta_{+} \cdot \Delta_{-}}\over {\Delta_{+} \cdot {\dot x}(\tau_{+})
 \, \Delta_{-} \cdot {\dot x}(\tau_{-})}}\, \Big[ {\cal E}
 + (\tau_{+} - \tau_{-})\, {{\partial {\cal E}}\over {\partial
 \tau_{+}}}\Big]\nonumber \\
 &&\Big[ 1 - {\cal E} + (\tau_{+} - \tau_{-})\,  {{\partial
 {\cal E}}\over {\partial \tau_{-}}} \Big] +
 (\tau_{+} - \tau_{-})^2\, \Big({{\partial {\cal E}}\over {\partial\,
 {\hat n}_{(\tau_{-})} }}\Big)^2\, \Big({{\partial {\hat n}_{(\tau_{-})}}\over
 {\partial z^{\mu}}}\Big)^2 +\nonumber \\
 &+& 2\, (\tau_{+} - \tau_{-})\, \Big[ {\cal E}
 + (\tau_{+} - \tau_{-})\, {{\partial {\cal E}}\over {\partial
 \tau_{+}}}\Big]\, {{\partial {\cal E}}\over {\partial\,
 {\hat n}_{(\tau_{-})} }}\,
{{\triangle_{+} \cdot {{\partial {\hat n}_{(\tau_{-})}}\over
 {\partial z^{\mu}}}}\over {\triangle_{+} \cdot {\dot x}(\tau_{+})}}
 > 0,\nonumber \\
 &&{}\nonumber \\
 &&\nonumber\\
  &&\quad for\, every\,\,\, \tau_{-}, \theta_{(\tau_{-})}, \phi_{(\tau_{-})},
 \tau_{+}.
 \label{VI14}
 \eea

\noindent {\it This is the condition on the function ${\cal
E}(\tau_{-}, {\hat n}_{(\tau_{-})}, \tau_{+})$ to have an
admissible foliation}.
\medskip

Since $ \sgn\, {{\Delta_{+} \cdot \Delta_{-}}\over {\Delta_{+}
\cdot {\dot x}(\tau_{+})
 \, \Delta_{-} \cdot {\dot x}(\tau_{-})}} > 0$, in the special case
 ${{\partial {\cal E}}\over {\partial {\hat n}_{(\tau_{-})}}} = 0$ it must be

 \bea
 &&\Big[ {\cal E}
 + (\tau_{+} - \tau_{-})\, {{\partial {\cal E}}\over {\partial
 \tau_{+}}}\Big]\,
 \Big[ 1 - {\cal E} + (\tau_{+} - \tau_{-})\,  {{\partial
 {\cal E}}\over {\partial \tau_{-}}} \Big] > 0,\nonumber \\
 &&{}\nonumber \\
 &&\Downarrow\nonumber \\
 &&{}\nonumber \\
 && {\cal E} + (\tau_{+} - \tau_{-})\, {{\partial {\cal E}}\over {\partial
 \tau_{+}}} \lessgtr 0,\qquad
 1 - {\cal E} + (\tau_{+} - \tau_{-})\,  {{\partial
 {\cal E}}\over {\partial \tau_{-}}}  \lessgtr 0.
 \label{VI15}
 \eea

\medskip

If $x^{\mu} = z^{\mu}(\tau ,\vec \sigma )$ is the admissible
embedding generating the four functions ${\cal E}$ and ${\vec
{\cal G}}$, the inverse transformation $x^{\mu}\, \mapsto\,
\Big(\tau (x); \vec \sigma (x) = {\vec {\cal G}}(x)\Big)$ allows
to define the inverse metric $g^{AB}(\tau ,\vec \sigma ) =
{{\partial \sigma^A(x)}\over {\partial x^{\mu}}}\, {{\partial
\sigma^B(x)}\over {\partial x^{\nu}}}\, \eta^{\mu\nu}$, also
satisfying Eqs.(\ref{I1}),  and the condition (\ref{VI14}) on
${\cal E}$ turns out to be nothing else that $\sgn\,
g^{\tau\tau}(\tau ,\vec \sigma ) > 0$. As a consequence the four
functions ${\cal E}$ and ${\vec {\cal G}}$ generating an
admissible 3+1 splitting of Minkowski space-time must be such that
${\cal E}$ satisfies Eq.(\ref{VI14}) and ${\vec {\cal G}}$
generates an inverse 3-metric $g^{rs}(\tau ,\vec  \sigma ) =
{{\partial {\cal G}^r(x)}\over {\partial x^{\mu}}}\, {{\partial
{\cal G}^s(x)}\over {\partial x^{\nu}}}\, \eta^{\mu\nu}$ which
satisfies the conditions

\bea
 \sgn\, g^{rr}(\tau ,\vec \sigma ) < 0,&&
 \begin{array}{|ll|} g^{rr}
 & g^{rs} \\ g^{sr} & g^{ss} \end{array}\, (\tau ,\vec \sigma )\, > 0,
\qquad \sgn\, det\, |g^{rs}(\tau ,\vec \sigma )| < 0,\nonumber \\
 &&{}\nonumber \\
 \Rightarrow&& det\, |g^{AB}(\tau ,\vec \sigma )| < 0.
 \label{VI16}
 \eea

 Moreover, for $\sigma \rightarrow \infty$ the quantities
 $\Big({{\partial \tau (x)}\over {\partial x^{\mu}}}; {{\partial
 {\cal G}^r(x)}\over {\partial x^{\mu}}}\Big)$ must tend to a
 constant limit $\epsilon^A_{\mu} = (\epsilon^{\tau}_{\mu};
 \epsilon^r_{\mu})$, where the asymptotic cotetrad
 $\epsilon^A_{\mu}$ is dual to the asymptotic tetrad
 $\epsilon^{\mu}_A$ appearing in the embedding of the leaves of
 the admissible 3+1 splitting.

\medskip

Therefore, given the world-line $x^{\mu}(\tau )$ of an observer
$\gamma$ and four functions $0 < {\cal E}(\tau_{-}, {\hat
n}_{(\tau_{-})}, \tau_{+}) <1$ and ${\vec {\cal G}}(\tau_{-},
{\hat n}_{(\tau_{-})}, \tau_{+})\, {\rightarrow}_{\tau_{+}
\rightarrow \tau_{-}}\, 0$, with a finite limit for $\tau_{\pm}\,
\rightarrow \pm \infty$, with ${\cal E}(\tau_{-}, {\hat
n}_{(\tau_{-})}, \tau_{+})$ satisfying Eq.(\ref{VI14}) and with
${\vec {\cal G}}(\tau_{-}, {\hat n}_{(\tau_{-})}, \tau_{+})$
satisfying Eqs.(\ref{VI16}), we can build the admissible
observer-dependent 4-coordinates $\tau$, $\vec \sigma $ of a
$\gamma$-dependent notion of simultaneity, because
Eqs.(\ref{VI14}) and (\ref{VI16}) ensure that the surfaces
$\Sigma_{\tau}$ are the space-like leaves of an admissible 3+1
splitting.

\bigskip

As a consequence, to give four admissible functions ${\cal
E}(\tau_{-}, {\hat n}_{(\tau_{-})}, \tau_{+})$, ${\vec {\cal
G}}(\tau_{-}, {\hat n}_{(\tau_{-})}, \tau_{+})$ is equivalent to
define {\it an operational method to build a grid of radar
4-coordinates} associated with the arbitrarily given time-like
world-line $x^{\mu}(\tau ) = z^{\mu}(\tau ,\vec 0)$ of the
spacecraft $\gamma$, i.e. of an accelerated observer. The
reconstruction of the admissible embedding $z^{\mu} (\tau ,\vec
\sigma )$ (we have used Eq.(\ref{IV1}) as an example) is done
locally \cite{22a} by the observer $\gamma$ with a suitable
computer software, which, starting from the four functions, builds
the associated simultaneity surfaces on which the clocks of the
spacecrafts $\gamma_i$ are synchronized. This procedure simulates
the use of light signals emitted by $\gamma$ and reflected towards
$\gamma$ from the other spacecrafts $\gamma_i$. This justify the
name radar 4-coordinates.

\bigskip

In general relativity on globally hyperbolic space-times, we can
define in a similar way the admissible, dynamically determined
\cite{31,36}, global notions of simultaneity and the admissible
one-way velocities of test light. Then the knowledge of the
functions ${\cal E}$ and ${\vec {\cal G}}$, associated to an
admissible notion of simultaneity, will allow an operational
determination of the 4-coordinates $(\tau ,\vec \sigma )$ adapted
to the chosen notion of simultaneity with simultaneity surfaces
$\tau = const.$ as radar coordinates. This is a step towards
implementing the operational definition of space-time proposed in
Refs. \cite{35,36}. The lacking ingredient is an operational
confrontation of the tetrads ${}_{(\gamma_i)}E^{\mu}_A(\tau )$
with the tetrad ${}_{(\gamma)}E^{\mu}_A(\tau )$ of the reference
world-line: this would allow a determination of the 4-metric in
the built radar 4-coordinates on a finite region of space-time
around the $N + 1$ spacecrafts of the GPS type, whose trajectories
are supposed known (for instance determined with the standard
techniques of space navigation \cite{17} controlled by a station
on the Earth). See Refs.\cite{44} for other approaches to GPS type
coordinates.

\vfill\eject

\section{Conclusions.}

Traditionally, due to the relativity principle, special relativity
is presented in inertial frames and the role of the instantaneous
3-space of every inertial observer is taken by the $x^o = const.$
hyper-planes, on which all the clocks are assumed to be
synchronized. This choice implies the orthogonality of the
simultaneity surfaces to the straight world-lines of the inertial
observers and Einstein ${1\over 2}$ convention for the
synchronization of distant clocks.
\medskip

However, actual observers are always non inertial and rotating.
Till now all the efforts to define an instantaneous 3-space for an
accelerated observer tried to preserve the two properties of
orthogonality of the simultaneity surfaces to the observer
world-line and of Einstein synchronization. The consequence of
these requirements was the appearance of coordinate singularities:
Fermi coordinates are only a local coordinate chart defined in a
suitable world-tube around the accelerated observer. Analogously,
in extended rotating relativistic systems like the rotating disk
there is no accord on which can be the disk instantaneous 3-space
and the use of Einstein convention, starting from the rotation
axis, leads to synchronization gaps and discontinuities (see
Ref.\cite{15} and its bibliography). Again there are coordinate
singularities signalled by the pathologies of the rotating
4-metrics.

With only a local coordinate chart instead of an equal-time Cauchy
surface with a good atlas of coordinates, the accelerated observer
cannot integrate Maxwell equations and check the validity of the
conservation laws.

\medskip

These problems become worse in general relativity, even in the
case of globally hyperbolic space-times. Fermi or Martzke-Wheeler
4-coordinates again constitute only local coordinate charts around
the (either geodesic or accelerated) world-line of a time-like
observer and the use of Einstein convention \cite{23a} for the
definition of an instantaneous 3-space (identified with a tangent
space) has a limited range of validity.
\bigskip

In this paper we have emphasized that, given an arbitrary
accelerated observer, the definition of an instantaneous 3-space
is both {\it observer-dependent and conventional} and that there
are as many possibilities as admissible 3+1 splittings of
space-time, namely of nice foliations with space-like
hyper-surfaces satisfying M$\o$ller and asymptotic (at spatial
infinity) admissibility conditions, implying for instance the {\it
non-existence of relativistic rigidly rotating frames} (only
differential rotations are admissible). Each admissible 3+1
splitting allows to find an observer-adapted globally defined
radar 4-coordinate system for non-inertial frames centered on
anyone of the time-like observers of the two naturally associated
congruences and a globally defined notion of instantaneous 3-space
which is also a good Cauchy surface for the equations of motion.
Besides the Eulerian observers with world-lines orthogonal to the
simultaneity surfaces, each admissible 3+1 splitting identifies a
non-orthogonal congruence of time-like observers with non-zero
vorticity, whose unit 4-velocity field (the evolution vector field
of the foliation) simulates a rotating disk. In this way a
foliation, whose leaves are genuine instantaneous 3-spaces for a
rotating disk, may be identified with standard notions  of spatial
distances and one-way velocity of light based on a well defined
modification of Einstein convention.
\medskip

These notions of instantaneous 3-space for an accelerated observer
are globally defined and  the synchronization of distant clocks is
done with a generalization of Einstein convention (following an
old suggestion of Havas \cite{30}). We have also suggested an
empirical method for a local construction of this type of radar
coordinates around an accelerated observer (a spacecraft of a
cluster like in GPS).

\medskip

Finally, when we have a Lagrangian description of an isolated
system in special relativity, it can be generalized to a
parametrized Minkowski theory in which the embeddings of the
simultaneity surfaces are {\it configuration gauge variables}. The
restricted general covariance (frame-preserving diffeomorphisms)
of these theories implies that all the admissible conventions
about the instantaneous 3-space, to be also used as Cauchy surface
for the equations of motion of the system, are {\it gauge
equivalent}. Each gauge choice identifies a non-rigid non-inertial
frame (i.e. a physical extended non-inertial laboratory)) with a
notion of instantaneous 3-space and the gauge transformations are
nothing else that the coordinate transformations among such
frames.
\bigskip

In general relativity in globally hyperbolic, asymptotically flat
at spatial infinity space-times without super-translations, we
have the same pattern if we use the Hamiltonian treatment of
metric and tetrad gravity developed in Refs.\cite{35,36}. As shown
in Refs.\cite{1,30,15} the definition of an instantaneous 3-space
is again observer-dependent and conventional and all the
admissible conventions (3+1 splittings) are gauge equivalent. But
now, since the chrono-geometrical structure of a general
relativistic space-time is dynamical, each solution of Hamilton
equations (i.e. of Einstein's equations) identifies on-shell a
dynamical notion of simultaneity in each coordinate system
admissible for that solution (see Refs.\cite{31,36} for more
details). In other words, in each Einstein's space-time there is a
dynamical emergence of a notion of instantaneous 3-space in each
coordinate system, namely a dynamical definition of the non-rigid
non-inertial frame using that coordinate system (i.e. a physical
extended laboratory) \cite{24a}.
\bigskip

Till now all these problems have been considered {\it academic}
and all space experiments around the Earth (GPS included) have
been done by replacing the conventions for the choice of an
instantaneous 3-space with a set of empirical (often
semi-Galilean) transformation rules and with ad hoc Sagnac
corrections of the rotating clocks to match descriptions in
different accelerated local coordinate systems (one of them is
always a rotating Earth-fixed one).
\medskip

However, the development of high precision laser cooled clocks and
their synchronization with one-way light signals \cite{16}, space
navigation in the solar system \cite{17}, GPS and Galileo system
\cite{18}, VLBI \cite{47}, LISA \cite{21} are pointing towards the
necessity to rephrase relativistic phenomena in the 3+1 framework
of this paper by using conventional globally defined
observer-dependent radar coordinates not influenced by the
rotation of the Earth. They would correspond to physical space
laboratories (non-rigid non-inertial reference frames) with the
transformation rules among them given by suitable frame preserving
diffeomorphisms. In the post-Newtonian approximation to the
gravitational field of the geoid these laboratories are further
restricted by the requirement of reproducing the (dynamically
determined) post-Newtonian 4-metric around the geoid.

\bigskip

Let us finish with a list of problems to be treated in the  next
future:

1) Write explicitly the coordinate transformations between the
admissible radar coordinate systems used in Section VID of
Ref.\cite{15} for the one-way time delay of the signals between
Earth and a satellite and the (locally defined) rotating
laboratories fixed on the Earth.

2) Since the observer-dependent radar coordinates are Lorentz
scalars, study how the phenomena of length contraction and time
dilation (evident in inertial Cartesian coordinates) are
reformulated in this language.

3) Include the (dynamically determined) post-Newtonian effects of
the gravitational field of the solar system, like it happens with
the prescribed IAU 4-metrics for the Solar and Geocentric
Celestial Frames \cite{2}, which however have fixed axes
identified by the fixed stars like in Minkowski space-time.

$\quad$ 3A) If we insist that these 4-metrics live in Minkowski
space-time, find the post-Newtonian M$\o$ller admissible
simultaneity surfaces (they cannot be hyper-planes $x^o =
const.$).

$\quad$ 3B) If the IAU frames live in general relativistic
globally hyperbolic space-times in which the Sun or the Earth are
accelerated observers, find which theoretical type (harmonic,
3-orthogonal, ..) of 4-coordinates is approximated by the IAU
4-metrics and fixed axes and which are the dynamically allowed
simultaneity surfaces.

4) In particular, try to understand to which theoretical type of
4-coordinates are associated the empirical NASA coordinates used
for the orbits of satellites \cite{48}. We have to identify some
experiment, whose post-Newtonian description in different
theoretical coordinates is possible, to be used to make a {\it
theoretical calibration} of the empirical coordinates. For
instance the (coordinate- and fixed star- dependent) Shapiro time
delay of the geoid to be measured by the future  ESA ACES mission
on the synchronization of laser cooled clocks located on the Earth
and on the space station \cite{16}. Also the synchronization of
the three clocks on the LISA spacecrafts will require the
framework developed in this paper.

5) Try to translate the astronomical, astrophysical and
cosmological definitions of {\it distance} of stars and galaxies
from the Earth in a notion of simultaneity, i.e. in a convention
on the synchronization of distant clocks, since, differently from
experiments inside the solar system with spacecrafts with
synchronized clocks, here the clock on the star is replaced by a
definition of distance and a hypothesis on the one-way velocity of
light. Do all these distances correspond to Einstein ${1\over 2}$
convention, like it is probably true for the parallax distance?

\vfill\eject

\appendix

\section{Notion of Simultaneity Associated to Rotating Reference
Frames.}

In this Section we consider the inverse problem of finding a
foliation of Minkowski space-time with simultaneity surfaces
associated to a given  arbitrary reference frame with non-zero
vorticity, namely to a time-like vector field whose expression in
Cartesian 4-coordinates in an inertial system is ${\tilde
u}^{\mu}(x)$ with ${\tilde u}^2(x) = \sgn$. In other words we are
looking for embeddings $z^{\mu}(\tau ,\vec \sigma )$, inverse of
an admissible 4-coordinate transformation, such that we have
${\tilde u}^{\mu}(z(\tau,\vec{\sigma})) = u^{\mu}(\tau ,\vec
\sigma ) = z^{\mu}_{\tau}(\tau ,\vec \sigma )/\sqrt{\sgn\,
g_{\tau\tau}(\tau ,\vec \sigma )}$. Let us remark that if the
vorticity is zero, the vector field ${\tilde u}^{\mu}(x)$ is
surface-forming, there is  a foliation whose surfaces have the
normal field proportional to $u^\mu(\tau,\vec{\sigma})$ and these
surfaces automatically give an admissible foliation with
space-like hyper-surfaces of Minkowski space-time.

\bigskip

Let us first show that, given an arbitrary time-like vector field
${\tilde u}^{\mu}(x)$, the looked for foliation exists. Let us
consider the equation

\beq
 {\tilde u}^{\mu}(x) \, {{\partial s(x)}\over {\partial x^{\mu}}}
 = 0,
 \label{V1}
 \eeq

\noindent where $s(x)$ is a scalar field. This equation means that
$s(x)$ is constant along the integral lines $x^{\mu}(s)$ [$d
x^{\mu}(s)/ {ds} = {\tilde u}^{\mu}(x(s))$] of the vector field,
i.e. it is a comoving quantity, since

\beq
 {{d s(x(s))}\over {ds}} = {\tilde u}^{\mu}(x(s))\, {{\partial
s}\over {\partial x^{\mu}}}(x(s)) = 0.
 \label{V2}
  \eeq

Since Eq.(\ref{V1}) has three independent solutions $s^{(r)}(x)$ ,
$r=1,2,3$, they can be used to identify three coordinates $
\sigma^r(x) = s^{(r)}(x)$. Moreover the three 4-vectors
${{\partial \sigma^r(x)}\over {\partial x^{\mu}}}$ are space-like
by construction.
\bigskip

Since Minkowski space-time is globally hyperbolic, there exist
{\it time-functions} $\tau (x)$ such that i) $\tau (x) = const.$
defines space-like hyper-surfaces; ii) ${{\partial \tau (x)}\over
{\partial x^{\mu}}}$ is a time-like 4-vector.
\medskip

As a consequence we can build an invertible 4-coordinate
transformation $x^{\mu} \mapsto \sigma^A(x) = (\tau (x),
\sigma^r(x))$, with inverse $\sigma^A = (\tau , \sigma^r) \mapsto
x^{\mu} = z^{\mu}(\tau ,\vec \sigma )$ for every choice of $\tau
(x)$. It can be shown that we get always a non-vanishing Jacobian
\cite{25a}

\beq
 J = \det\, \Big( {{\partial \tau (x)}\over {\partial
x^{\mu}}}, {{\partial \sigma^r(x)}\over {\partial x^{\mu}}}\Big)
\not= 0. \label{V3} \eeq

By using

\beq {{\partial \sigma^A(x)}\over {\partial x^{\nu}}}\, {{\partial
x^{\mu}}\over {\partial \sigma^A}}(\sigma (x)) = \eta^{\mu}_{\nu},
 \label{V4}
 \eeq

\noindent and Eq.(\ref{V1}) we get the desired result

\beq
 {\tilde u}^{\mu}(x) = {\tilde u}^{\nu}(x)\, {{\partial
\sigma^A(x)}\over {\partial x^{\nu}}}\, {{\partial x^{\mu}}\over
{\partial \sigma^A}}(\sigma (x)) = \Big( {\tilde u}^{\nu}(x)\,
{{\partial \tau (x)}\over {\partial x^{\nu}}}\Big)\, {{\partial
z^{\mu}(\tau ,\vec \sigma )}\over {\partial \tau}} =
{{z^{\mu}_{\tau}(\tau ,\vec \sigma )}\over {\sqrt{\sgn \,
g_{\tau\tau}(\tau ,\vec \sigma )}}}.
 \label{V55}
  \eeq

\bigskip

Given a unit time-like vector field ${\tilde u}^{\mu}(x) =
u^{\mu}(\tau ,\vec \sigma )$ such that $u^{\mu}(\tau ,\vec \sigma
)\, {\rightarrow}_{|\vec \sigma | \rightarrow \infty}\,
n^{\mu}(\tau )$ and ${{\partial u^{\mu}(\tau ,\vec \sigma )}\over
{\partial \sigma^r}}\,   {\rightarrow}_{|\vec \sigma | \rightarrow
\infty}\, 0$, to find the embeddings $z^{\mu}(\tau ,\vec \sigma )$
we must integrate the equation

\beq
 {{\partial z^{\mu}(\tau ,\vec \sigma )}\over {\partial
\tau}} = f(\tau ,\vec \sigma )\, u^{\mu}(\tau ,\vec \sigma
),\qquad u^2(\tau ,\vec \sigma ) = \sgn,
 \label{V6}
 \eeq

\noindent where $f(\tau ,\vec \sigma )$ is an integrating factor.
\medskip

Since Eq.(\ref{V6}) implies $\sgn \, g_{\tau\tau}(\tau ,\vec
\sigma ) = f^2(\tau ,\vec \sigma ) > 0$, the only restrictions on
the integrating factor are:

i) it must never vanish;

ii) $f(\tau ,\vec \sigma )\,  {\rightarrow}_{|\vec \sigma |
\rightarrow \infty}\, f(\tau )$ finite.
\bigskip

The integration of Eq.(\ref{V2}) gives

\begin{eqnarray*}
 z^{\mu}(\tau ,\vec \sigma ) &=& g^{\mu}(\vec \sigma ) +
 \int_o^{\tau} d\tau_1\, f(\tau_1, \vec \sigma )\, u^{\mu}(\tau_1,
 \vec \sigma ),\nonumber \\
 &&{}\nonumber \\
 &&\Downarrow\nonumber \\
 &&{}\nonumber \\
 z^{\mu}_r(\tau ,\vec \sigma )&=& \partial_r\, g(\vec \sigma ) +
 \int_o^{\tau} d\tau_1\, \partial_r\, [f(\tau_1, \vec \sigma )\,
 u^{\mu}(\tau ,\vec \sigma )],
 \end{eqnarray*}

\bea
 g_{\tau r}(\tau ,\vec \sigma ) &=&f(\tau ,\vec \sigma )\,
 u_{\mu}(\tau ,\vec \sigma )\, \Big[ \partial_r\, g(\vec \sigma ) +
 \int_o^{\tau} d\tau_1\, \partial_r\, [f(\tau_1, \vec \sigma )\,
 u^{\mu}(\tau ,\vec \sigma )]\Big]\nonumber \\
&&\nonumber\\
 && {\rightarrow}_{|\vec \sigma | \rightarrow \infty}\,
 f(\tau )\, n_{\mu}(\tau )\, \Big[ \lim_{|\vec \sigma | \rightarrow \infty}\,
\partial_r\, g(\vec \sigma )\Big],
 \label{V7}
 \eea

\noindent where $g(\vec \sigma )$ is arbitrary and we have assumed
that the integrating factor satisfies $\partial_r\, f(\tau ,\vec
\sigma )\, {\rightarrow}_{|\vec \sigma | \rightarrow \infty}\, 0$.
\bigskip

For the sake of simplicity let us choose $g(\vec \sigma ) =
\epsilon^{\mu}_r\, \sigma^r$ with the constant 4-vectors
$\epsilon^{\mu}_r$ belonging to an orthonormal tetrad
$\epsilon^{\mu}_A$. Then $g_{\tau r}(\tau ,\vec \sigma )$ has the
finite limit $f(\tau )\, n_{\mu}(\tau )\, \epsilon^{\mu}_r$.

With this choice for $g(\vec \sigma )$ we get

\bea
 z^{\mu}_r(\tau ,\vec \sigma ) &=& [\delta_{rs} + \alpha_{rs}(\tau
 ,\vec \sigma )]\, \epsilon^{\mu}_s + \beta_r(\tau ,\vec \sigma
 )\, \epsilon^{\mu}_{\tau},\nonumber \\
 &&{}\nonumber \\
 &&\quad \alpha_{rs}(\tau ,\vec \sigma ) =
 \int_o^{\tau} d\tau_1\, \partial_r\, [f(\tau_1 ,\vec \sigma )\,
 \epsilon_{s \mu}\, u^{\mu}(\tau_1, \vec \sigma )],\nonumber \\
&&\nonumber\\
 &&\quad \beta_r(\tau ,\vec \sigma ) =
 \int_o^{\tau} d\tau_1\, \partial_r\, [f(\tau_1 ,\vec \sigma )\,
 \epsilon_{\tau \mu}\, u^{\mu}(\tau_1, \vec \sigma )].
 \label{V8}
 \eea
 \medskip

 Since $u^{\mu}(\tau ,\vec \sigma )$ and $\epsilon^{\mu}_{\tau}$
 are future time-like [$\sgn\, u^o(\tau ,\vec \sigma ) > 0$,
  $\sgn\, \epsilon^o_{\tau} > 0$], we have $u^{\mu}(\tau ,\vec \sigma ) =
 \sgn\, a(\tau ,\vec \sigma )\, \epsilon^{\mu}_{\tau} +
 b_r(\tau ,\vec \sigma )\, \epsilon^{\mu}_r$ with $a(\tau ,\vec
 \sigma ) > 0$ and without zeroes.
\bigskip

 Let us determine the integrating factor $f(\tau ,\vec \sigma )$
 by requiring $\beta_r(\tau ,\vec \sigma ) = 0$ as a consequence
 of the equation

 \begin{eqnarray*}
 0 &=& \sgn\, \partial_r\, [f(\tau ,\vec \sigma )\, \epsilon_{\tau \mu}\,
 u^{\mu}(\tau ,\vec \sigma )] = f(\tau ,\vec \sigma )\,
 \partial_r\, a(\tau ,\vec \sigma ) + \partial_r\, f(\tau ,\vec
 \sigma )\, a(\tau ,\vec \sigma ),\nonumber \\
 &&{}\nonumber \\
 &&\Downarrow\nonumber \\
 &&{}\nonumber \\
 f(\tau ,\vec \sigma ) &=& e^{c(\tau )}\, a(\tau ,\vec \sigma
 ),\end{eqnarray*}

\bea
 z^{\mu}_r(\tau ,\vec \sigma ) &=& [\delta_{rs} + \alpha_{rs}(\tau
 ,\vec \sigma )]\, \epsilon^{\mu}_s,\nonumber \\
&&\nonumber\\
 \alpha_{rs}(\tau ,\vec \sigma ) &=& \int_o^{\tau} d\tau_1\,
 e^{c(\tau_1)}\, \partial_r\, [a(\tau_1,\vec \sigma )\,
 \epsilon_{s \mu}\, u^{\mu}(\tau_1, \vec \sigma )],\nonumber \\
 &&{}\nonumber \\
 g_{rs}(\tau ,\vec \sigma ) &=& -\sgn\, \Big(\delta_{rs} +
 \alpha_{rs}(\tau ,\vec \sigma ) +  \alpha_{sr}(\tau ,\vec \sigma )
 + \sum_u\,  \alpha_{ru}(\tau ,\vec \sigma )\,  \alpha_{su}(\tau ,\vec \sigma )
\Big).
 \label{V9}
 \eea
\medskip

 Let us choose the arbitrary function $C(\tau ) = e^{c(\tau )}$ so
 small that $|  \alpha_{rs}(\tau ,\vec \sigma )| << 1$ for every
 $r$, $s$, $\tau$, $\vec \sigma$, so that all the conditions on
 $g_{rs}(\tau ,\vec \sigma )$ from Eqs.(\ref{I1}) are satisfied.
 \bigskip

 In conclusion given an arbitrary congruence of time-like
 world-lines, described by a vector field ${\tilde u}^{\mu}(x)$,
 an embedding defining a good notion of simultaneity is
 [$x^{\mu}(\tau ) \, {\buildrel {def}\over =}\, z^{\mu}(\tau ,\vec 0)$]

 \bea
  z^{\mu}(\tau ,\vec \sigma ) &=& \epsilon^{\mu}_r\, \sigma^r +
  \int_o^{\tau} d\tau_1\, C(\tau_1)\, \epsilon_{\tau \nu}\,
  u^{\nu}(\tau_1, \vec \sigma )\, u^{\mu}(\tau_1, \vec \sigma )
  =\nonumber \\
  &=& x^{\mu}(\tau ) +  \epsilon^{\mu}_r\, \sigma^r +
  \int_o^{\tau} d\tau_1\, C(\tau_1)\, \epsilon_{\tau \nu}\,
  \Big[u^{\nu}(\tau_1, \vec \sigma )\, u^{\mu}(\tau_1, \vec \sigma )
  - u^{\nu}(\tau ,\vec 0)\, u^{\mu}(\tau ,\vec 0)\Big],\nonumber  \\
 &&{}
  \label{V10}
  \eea

\noindent for sufficiently small $C(\tau )$. Here
$\epsilon^{\mu}_A$ is an arbitrary orthonormal tetrad.

\medskip

As a consequence, given any congruence associated to a rotating
disk, we can find admissible 3+1 splittings of Minkowskki
space-time, with the space-like simultaneity  leaves not
orthogonal to the rotation axis  of the disk, which allow to
define genuine instantaneous 3-spaces with synchronized clocks for
every rotating disk. See Ref.\cite{15} (Sections VIB and C) for
the 3+1 treatment of the rotating disk and the Sagnac effect along
these lines.

\vfill\eject


\begin{thebibliography}{}
\bibitem{1a}The time-like axis is given by observer's unit 4-velocity,
while the spatial axes correspond to a conventional choice of the
orientation and transport of three mutually orthogonal gyroscopes.
The Minkowski metric has signature $\sgn\, (+---)$ with $\sgn =
\pm 1$ according to the particle physics ($\sgn = 1$) or general
relativity ($\sgn = - 1$) convention.
\bibitem{2a}For an inertial observer $x^o$ coincides with the proper
time $\tau$ of the clock on $\gamma$ since the inertial frame is
the rest frame.
\bibitem{7a}We use the vector notation $\vec \sigma$ for notational
simplicity.
\bibitem{8a}While this hypothesis (implying that rods and clocks must
be assumed not to be influenced by acceleration; see for instance
Ref.\cite{22}) is verified in Newtonian mechanics and in those
relativistic cases in which a phenomenon can be reduced to
point-like coincidences of classical point particles and light
rays (geometrical optic approximation), its validity is
questionable in presence of electro-magnetic waves. As emphasized
by Mashhoon \cite{6,7}, in this case we can trust the locality
hypothesis only when the wave-length $\lambda$ of the
electro-magnetic wave is much shorter of the acceleration length
${\cal L}$ of the observer, describing the degree of variation of
its state, i.e. when $\lambda << {\cal L}$. When $\lambda << {\cal
L}$ holds, so that the period of the wave satisfies
${{\lambda}\over c} << {{{\cal L}}\over c}$, the observer state
does not change appreciably on the time scale needed to detect a
few oscillations of the wave and to measure its frequency. Instead
in the case of the electro-magnetic waves radiated by an
accelerating charged particle with acceleration length ${\cal L}$,
we have $\lambda \approx {\cal L}$. In this case it is highly
problematic to consider the particle momentarily equivalent to an
identical comoving inertial particle. This fact is confirmed by
the causality problems (pre-acceleration, runaway solutions) of
the classical Abraham - Lorentz - Dirac equation of motion of the
particle (see for instance Ref.\cite{25}), which depend on the
time derivative of the acceleration.
\bibitem{9a}This convention was already criticized in
Refs.\cite{26,27,28,29,30}.
\bibitem{10a}They would contribute \cite{24} to the asymptotic line
element with the diverging terms $[A_i(y^o)\, y^i + B_{ij}(y^o)\,
y^i\, y^j]\, (dy^o)^2$ and $\epsilon_{ijk}\, \omega^j(y^o)\, y^k\,
dy^o\, dy^i$, respectively.
\bibitem{11a}Their inverse $z^A_{\mu}(\tau ,\vec \sigma )$ are flat cotetrad
fields over Minkowski space-time: $z^A_{\mu}\, g_{AB}\, z^B_{\nu}
= \eta_{\mu\nu}$.
\bibitem{12a}In Ref.\cite{30} Havas proposed the following two examples
(the second one is a time-dependent transformation) of
simultaneity foliations associated with admissible 4-coordinates,
i.e. whose 4-metric satisfies Eq.(\ref{I1}):
\begin{eqnarray*}
 1) &&y^o =x^o + {1\over {c^2}}\, f(x^i),\qquad y^i = x^i,\qquad
 \mbox{with inverse } x^o = y^o - {1\over {c^2}}\, f(y^i),\quad
  x^i = y^i,\nonumber \\
  &&{}\nonumber \\
 &&\mbox{and with associated 4-metric } \nonumber\\
&&\nonumber\\
&& g_{oo}(y) = - c^2,\quad g_{oi}(y) = {{\partial f(y)}\over
{\partial y^i}}\,\, (\mbox{ if }  g_{oi} = const.,\mbox{ then }
(g_{oi})^2 <
c^2),\nonumber \\
&&\nonumber\\
 &&g_{ij}(y) =\delta_{ij} - {1\over {c^2}}\,
{{\partial f}\over {\partial y^i}}\, {{\partial f}\over {\partial
y^j}};
 \end{eqnarray*}
 \begin{eqnarray*}
 2) &&y^o = x^o\, f^{-1}(x^i),\qquad y^i = x^i,\qquad
 \mbox{with inverse } x^o = y^o\,
f(y^i),\quad x^i = y^i,\qquad \nonumber\\
&&{}\nonumber \\
&&\mbox{and with associated
4-metric }\nonumber\\
&&\nonumber\\
&&
 g_{oo}(y) = - c^2\, f^2,\qquad g_{oi}(y) = - y^o\, c^2\,
  f\, {{\partial f}\over {\partial y^i}},\qquad
 g_{ij}(y) = \delta_{ij} - (y^o)^2\, c^2\, {{\partial f}\over
{\partial y^i}}\, {{\partial f}\over {\partial y^j}}.
 \end{eqnarray*}
Both of them are examples of admissible 4-coordinate systems not
interpretable in terms of comoving observers as required by the
locality hypothesis.
\bibitem{13a}A variant are the embeddings
\medskip
\begin{equation*}
 z^{\mu}(\tau ,\vec \sigma ) = x_o^{\mu}(\tau ) +
\Lambda^{\mu}{}_{\nu}(\tau ,\vec \sigma )\, \epsilon^{\nu}_A\,
\sigma^A \, {\rightarrow}_{|\vec \sigma | \rightarrow \infty}\,
x^{\mu}_o(\tau ) + \epsilon^{\mu}_{\tau}\, \tau +
\epsilon^{\mu}_r\, \sigma^r = x^{\mu}(\tau ) + \epsilon^{\mu}_r\,
\sigma^r,
 \end{equation*}
\medskip
\noindent with $x^{\mu}(\tau ) = x^{\mu}_o(\tau ) +
\epsilon^{\mu}_{\tau}\, \tau = \epsilon^{\mu}_{\tau}\, [\tau +
f(\tau )]$ an arbitrary time-like straight-line (an inertial
observer) not parametrized in terms of the proper time. The only
difference now is that the asymptotic hyper-planes are no more
uniformly spaced like in the case $x^{\mu}(\tau ) = x^{\mu}_o +
\epsilon^{\mu}_{\tau}\, \tau$ ($z^{\mu}_{\tau} =
\epsilon^{\mu}_{\tau}\, \mapsto\, z^{\mu}_{\tau}(\tau ) =
\epsilon^{\mu}_{\tau}\, [1 + \dot f(\tau )]$).
\bibitem{14a}The second non-inertial and non-surface-forming congruence
of observers associated with these embeddings, with vector field
$z^{\mu}_{\tau}(\tau ,\vec \sigma )/\sqrt{\sgn\, g_{\tau\tau}(\tau
,\vec \sigma )}$, asymptotically tends to the same asymptotic
inertial reference frame because $z^{\mu}_{\tau}(\tau ,\vec \sigma
)\, {\rightarrow}_{|\vec \sigma | \rightarrow \infty}\,
\epsilon^{\mu}_{\tau} = l^{\mu}_{(\infty )}$ if $\partial_{\tau}\,
\Lambda^{\mu}{}_{\nu}(\tau ,\vec \sigma )\, {\rightarrow}_{|\vec
\sigma | \rightarrow \infty}\, 0$.
\bibitem{15a}$g_{oo} = \sgn ([(1+ {{\vec a \cdot \vec x}\over {c^2}})^2
- {{(\omega \times \vec x)^2}\over {c^2}})$, $g_{oi} = -\sgn\,
{1\over c}\, (\vec \omega \times \vec x)^i$, $g_{ij} = -\sgn\,
\delta_{ij}$, where $\vec a$ is the time-dependent acceleration of
the observer's frame of reference relative to the comoving
inertial frame and $\vec \omega$ is the time-dependent angular
velocity of the observer's spatial rotation with respect to the
comoving frame; $\vec x$ is the position vector of a spatial point
with respect to the origin of the observer's accelerated frame.
\bibitem{17a}This is the case of a (non-time-like) Rindler observer
with uniform 4-acceleration, see Ref.\cite{41}.
\bibitem{18a}As said in Subsections C and D
 of Section VI of Ref.\cite{15}, the choice $F(\sigma ) = {{1 +
 {{\omega^2\, R^2}\over {c^2}}}\over {1 + {{\omega^2\, \sigma^2}\over {c^2}}}}
 < {2\over {1 + {{\omega^2\, \sigma^2}\over {c^2}}}} <
 {1\over {\omega\, \sigma}}$, $F(\sigma )\, \rightarrow_{c\rightarrow
 \infty}\, 1 + {{\omega^2\, (R^2 - \sigma^2)}\over {c^2}} + O({1\over {c^4}})$,
 replaces the rigid rotation $\Omega (\sigma ) = \omega$ for $\sigma < R$ of a
 rotating disk of radius $R$ (with $\omega\, R < c$) with an admissible differential
 rotation $\Omega (\sigma ) = \omega\, F(\sigma )$. By varying the
 admissible functions $F(\sigma )$ (a gauge transformation in  parametrized
 Minkowski theories) we can approximate the step function $\Omega (\sigma ) = \omega$
 for $\sigma < R$, $\Omega (\sigma ) = 0$ for $\sigma > R$, as much as we
 wish without violating Eq.(\ref{IV14}).
\bibitem{19a}Let us remark that the conditions (2.1) may be replaced by
the single condition $\sgn\, g_{oo}(y) > 0$ plus the requirement
of the existence  of the inverse transformation, namely the
existence of a positive Jacobian.
\bibitem{20a}The emission unit 3-direction ${\hat n}_{(\tau_{-})}(
\theta_{(\tau_{-})} ,\phi_{(\tau_{-})} )$ could be replaced with a
reception unit 3-direction ${\hat n}_{(\tau_{+})}(
\theta_{(\tau_{+})} ,\phi_{(\tau_{+})} )$ if more convenient.
\bibitem{21a} If we introduce the function
\[
g_{\pm}(y)=x_l(y)-x_l(\tau ) \pm|{\vec x}_{\epsilon}(y)- \vec\xi
  (\tau ,\vec \sigma )|,
\]
Eqs. (\ref{VI10}) are equivalent to $g_{\pm}(y)=0$. The solution
is unique because the functions $g_{\pm}(y)$ are decreasing in
$y$, since we have
\[
\frac{dg_{\pm}(y)}{dy}=
-v_l(y)\pm\sum_r\,v_r(y)\,\frac{x^r_{\epsilon}(y)- \xi^r
  (\tau ,\vec \sigma )}{|{\vec x}_{\epsilon}(y)- \vec\xi
  (\tau ,\vec \sigma )|}.
\]
Using  Eq.(\ref{IV11}) in the form
\[
\sum_r\,v_r(y)\,\frac{x^r_{\epsilon}(y)- \xi^r
  (\tau ,\vec \sigma )}{|{\vec x}_{\epsilon}(y)- \vec\xi
  (\tau ,\vec \sigma )|}
\leq |\vec{v}(y)|<v_l(y),
\]
we get $ \frac{dg_{\pm}(y)}{dy}<0$, since $v_l(y)>0$.
\bibitem{23a}See for instance the Landau-Lifschiz 3-metric \cite{45}
and its criticism in Ref.\cite{46} and in Section II of
Ref.\cite{15}.
\bibitem{24a}In absence of matter the dynamical admissible 3+1 splittings of
Minkowski space-time considered as a special solution of Einstein
equations (and not as an autonomous theory, special relativity)
must have the simultaneity hyper-surfaces 3-conformally flat due
to the vanishing of the Cotton-York tensor \cite{24,35}. This
restriction does not exist in special relativity, which,
therefore, admits a bigger family of non-dynamical admissible
non-inertial frames.
\bibitem{25a}Let us show that the equations
\[
\alpha\,\frac{\partial\tau(x)}{\partial x^\mu}+\beta_r\,
\frac{\partial\sigma^r(x)}{\partial x^\mu}=0
\]
implies $\alpha=\beta_r=0$. If we multiply for $\tilde{u}^\mu(x)$,
we get $\alpha\,\tilde{u}^\mu(x)\,\frac{\partial\tau(x)}{\partial
x^\mu}=0$. But $\frac{\partial\tau(x)}{\partial x^\mu}$ and
$\tilde{u}^\mu(x)$ are both time-like with
$\tilde{u}^\mu(x)\,\frac{\partial\tau(x)}{\partial x^\mu}\neq 0$,
so that we get $\alpha=0$. We remain with the equations $\beta_r\,
\frac{\partial\sigma^r(x)}{\partial x^\mu}=0$, which imply
 $\beta_r=0$ since the
$\frac{\partial\sigma^r(x)}{\partial x^\mu}$ are independent by
construction.
\end{thebibliography}
\end{document}